\documentclass[a4paper,fleqn]{cas-sc}
\usepackage[numbers]{natbib}
\usepackage{float}
\usepackage{graphicx}
\graphicspath{figures/}
\usepackage[justification=centering]{caption}
\usepackage{subfig}
\usepackage{hyperref}
\usepackage{romannum}
\usepackage{array}
\usepackage{multirow}
\usepackage{lineno}

\begin{document}
\let\WriteBookmarks\relax
\def\floatpagepagefraction{1}
\def\textpagefraction{.001}

\shorttitle{Coaxial HPGe detector}
\shortauthors{Swati Thakur et~al.}  

\title [mode = title] {Spectroscopic performance evaluation and modeling of a low background HPGe detector using GEANT4}

\affiliation[1]{organization={Department of Physics, Indian Institute of Technology Ropar, Rupnagar - 140 001, Punjab, India}}

\author[1]{Swati Thakur}
\author[1]{Soni Devi}
\author[1]{Sanjeet S. Kaintura}
\author[1]{Katyayni Tiwari}
\author[1]{Pushpendra P. Singh} [orcid=0000-0003-3813-4322]
\cortext[1]{Pushpendra P. Singh}
\ead{pps@iitrpr.ac.in}

\begin{abstract}
Low background gamma spectrometry employing HPGe detectors is a sensitive technique for measuring low-level radioactivity in environmental applications, material screening, and for rare decay searches. This work presents spectroscopic performance evaluation and modelling of a
low background measurement setup developed at IIT Ropar in Punjab, India, to measure trace natural radioactive elements, with a particular interest in studying low-level radioactivity in soil and/or rock samples to generate specific inputs for low background experiments. The performance test and characterization of a low background cryocooled HPGe detector with relative efficiency of $\sim$33\% have been carried out. An effective detector model has been developed using GEANT4 Monte Carlo simulation to determine the response of the detector over an energy range of 80.9--1408~keV and compared with the experimental performance of the detector. The response of the detector obtained using Monte Carlo simulations agrees reasonably well within 93\% level of confidence, indicating only 7\% deviation in the comparison. The present setup offers improved detection limits of primordial radionuclides (U/Th and K) to measure radioactive contamination in environmental matrices, which has been used elsewhere~\cite{phdthesis}.
\end{abstract}

\begin{keywords}
Carbon loaded HPGe detector, \sep Detector scanning, and characterization, \sep GEANT4, \sep Monte Carlo simulation, \sep Gamma-ray spectroscopy, \sep Low background measurements \sep Environmental radioactivity, \sep Soil sample analysis
\end{keywords}

\maketitle

\section{Introduction}\label{Introduction}
Gamma spectroscopy setups employing low background HPGe spectrometers play an increasingly important role in measurements of environmental radioactivity and material selection for rare event experiments~\cite{Heusser1995,Arpesella2002,Laubenstein2004,Budjas2009,Kohler2009,dolinski2019neutrinoless,Laubenstein2020}. The experimental sensitivity of these low background measurements depends critically on the identification and minimization of confounding background. 
In above-ground measurements, the background is mainly dominated by cosmic rays and cosmic ray-induced interactions, cosmogenic radionuclides (T$_{1/2}$ $\sim$\,d--y), primordial radioactivity consisting of gamma-ray radiation from natural decay chains of $^{238}$U, $^{232}$Th and $^{40}$K (T$_{1/2}$ $\sim$\,$10^{8}$--$10^{10}$\,y) in and around the detector setup material, airborne radioactivity of radon and its progeny and the detector itself is the source of $\alpha$, $\beta$, $\gamma$ and neutrons background~\cite{dolinski2019neutrinoless}. The environmental radioactivity originating from naturally occurring radionuclides, namely; $^{238}$U, $^{232}$Th, and $^{40}$K are long-lived ($T_{1/2}$ $\sim$$10^{8}$ - $10^{10}$~y) and present in trace amounts in the earth crust. The gamma-ray background from these naturally occurring radionuclides is a significant source of radiation exposure for above-ground laboratories. Measurement at the environmental level deals with low radioactivity concentration and involves inherent complexity due to the typical interference of natural radionuclides over a wide energy range. Additionally, the gamma-ray flux is subject to variation, possibly due to experiment location in the laboratory/site, seasonal fluctuations, the radon concentration near the experimental setup, or unknown systematic uncertainties. Background statistical fluctuations can influence the assessment of peak identification, peak area calculations, energy resolution, confined intervals, and detection limits. These background fluctuations can cause differences in the precise determination of activity concentrations in weak samples. For reliable estimation of the radioactive impurities, it is necessary to identify and minimize the background to improve the minimum detection limits of the spectrometer. Due to the typical interference of natural radionuclides, it is essential to accurately estimate the background level by measuring the concentration of the members of these decay chains and their variation~\cite{Bossew2005,Lenka2013}.   

In India, the proposals for a potential underground laboratory driven by the interest in rare event studies have been initiated~\cite{mondal2012india,banik2021simulation,thakur2022radiopurity}. As mentioned earlier, other than the ambient background from the surroundings, the material of the counting detectors itself contributes to the background contamination levels in rare event experiments. To improve the sensitivity of measurement and accurately determine radio impurities, an essential prerequisite is low radioactivity around the detector as much as possible. The low background cryostats contribute miniature to the sample spectrum, improving its signal-to-noise ratio and significantly reducing the Minimum detectable activity (MDA). Therefore, the radiopure cryostat with selectively refined detector construction materials and applying passive shielding allows for lower minimum detectable activity and high sample throughput for a specific counting time, which is highly desirable in low background counting applications. Furthermore, cryocooler-based low background HPGe detectors are desirable for probing low background experiments in above-ground or remote underground locations for long-term counting measurements. To maximize the counting efficiency for investigating trace radioactive elements, it is necessary to understand the detector performance in compact geometry over a wide energy range and different counting setup configurations. 

With this motivation, we have taken up the initiative to build a low background measurement setup at the Indian Institute of Technology (IIT) Ropar using the HPGe detectors. This setup is intended for radio-purity assessment and understanding of the radiation background at IIT Ropar to carry out low background experiments. This paper describes the performance test and characterization of a low background carbon fiber HPGe detector with a relative efficiency (R.E.) of $\sim$33\%. The first set of studies presents detector characterization with radioactive sources over a wide energy range from 80.9--1408\,keV. The performance specifications are successfully tested against the warranted values provided by the manufacturer. Measurements have been performed with point-like sources to calculate the peak shape, Peak-to-Compton ratio (P/C), Full energy peak efficiency (FEPE), and Figure of merit (FOM). Mono-energetic sources are used to perform close scanning of Ge crystal along its parallel and perpendicular axis to determine the inner crystal structure. A Monte Carlo model has been developed using GEANT4 to calculate the FEPE of the HPGe detector for gamma spectrometry measurements associated with the development of a low background setup.
It has been observed that germanium crystal volume and dead layers in the surroundings are crucial parameters in the detector geometry~\cite{HultMin2019, EsraIn2022}. Over the operational lifespan of the detector, the continuous diffusion of lithium ions into the germanium crystal leads to an increase in the dead layer thickness and a decrease in the active volume of the detector. Previous literature has reported a significant deviation of 10$\%$--30$\%$ between the experimental and simulated FEPE efficiencies in the energy range of 60-2000~keV~\cite{EsraIn2022,Dokania2014,Aguilar2020,Alexandre2022}. In order to validate a simulation code, the detector has been optimized, which typically involves adjusting various detector parameters, such as the dimensions, dead layer thickness, and other relevant factors.
The second set of studies characterizes the natural gamma-ray background during long counting measurement and its further reduction by incorporating a Pb shield at the above-ground laboratory of IIT Ropar. The organized statistical data on this topic would be valuable for inter-laboratory comparisons and unique in different geological and geographical conditions than the rest of the studied areas, especially quantifying low-level natural radioactivity. The activity concentration and Minimum detectable activity (MDA) for common radioisotopes in environmental samples are evaluated. The presentation of this paper is organized as follows, section~\ref{Experimental setup and Methodology} describes the experimental setup and procedure details, section~\ref{Results and Analysis} discusses the experimental characterization, detector modelling, comparison with simulated data, and spectroscopy measurements for characterizing background distributions. The summary and conclusions are given in section~\ref{Summary and Conclusions}.

\section{Experimental setup and Methodology}\label{Experimental setup and Methodology}
\begin{figure}[!ht]
\centering
\captionsetup{justification=justified,font=sf,labelfont=bf}
\includegraphics[width=0.70\linewidth,height=0.30\linewidth]{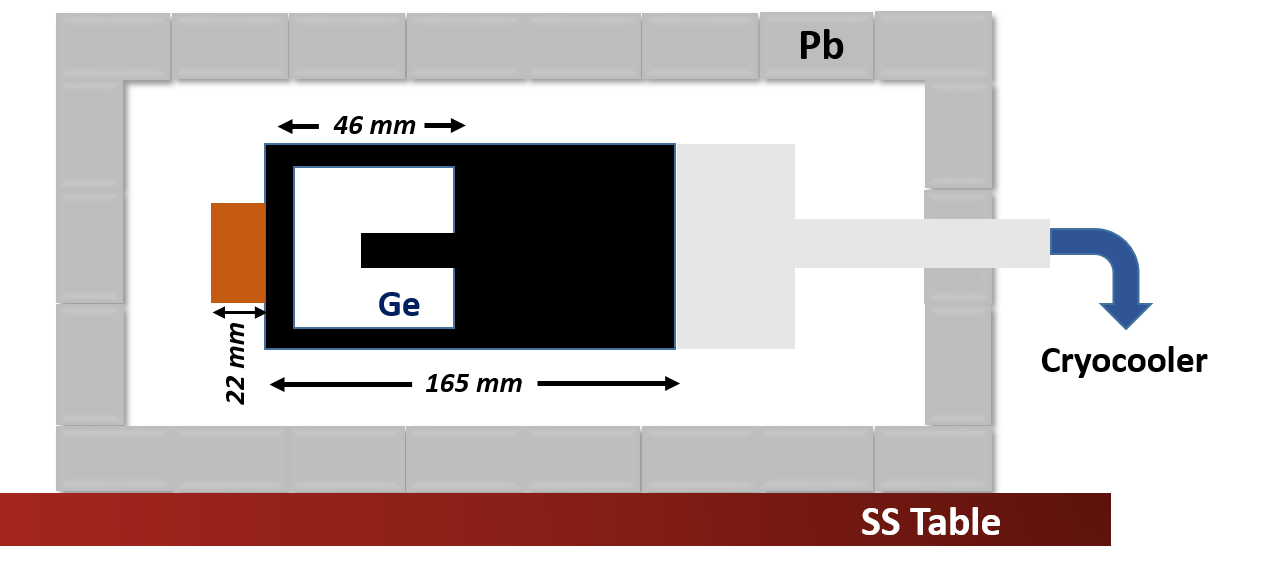}
\caption{Schematic cross-section of the experimental setup of ILM-0 (IIT Ropar Low Background Measurement setup). The soil sample is mounted on the face of the detector, and the setup is placed on a thick stainless steel (SS) table 1~m above the ground.}
\label{fig1}
\end{figure}

The experimental setup (ILM-0) comprises of p-type co-axial HPGe detector, ORTEC make (GEM30P4-83-RB) with a measured relative efficiency of 33\% having a crystal diameter of 62\,mm and thickness of 46\,mm. The detector crystal is mounted inside a cryostat of low background carbon fiber (carbon-composite) body, and the top face is 5\,mm beneath a 0.9\,mm thick carbon fiber entrance window. The rear end of the detector is attached to the $\sim$\,150\,mm long vertical cold finger connected to the transfer hose of an electro-mechanical cooler (ORTEC X-Cooler III) to cool the detector crystal to liquid nitrogen (LN$_2$) temperature (77~K). The typical cooldown time required by the electro-mechanical cooler is found to be 10-16~h. The detector assembly is supported with additional online UPS to protect during utility power failure. 
The HPGe detector is surrounded by moderate passive shielding with lead bricks of dimensions $22.9~\textrm{cm} \times7.6~\textrm{cm} \times5.1~\textrm{cm}$ and mounted on a custom-made stainless steel (SS) table, as shown in Figure~\ref{fig1}. In this arrangement, the detector with a preamplifier is entirely shielded with adequate space for mounting voluminous samples generally used in environmental radioactivity measurements. In addition, it has a provision for an inner layer of passive shielding and an active muon veto. It should be mentioned that different shapes and configurations of the lead shields were also considered before the final implementation of bricks geometry. 
The detector is coupled with the pulse processing electronics and data acquisition system, including a NIM-based high voltage power supply, spectroscopic amplifier, and multichannel analyzer procured from ORTEC. The multichannel analyzer records the data using the emulator software Maestro. A list of relevant technical details, along with the associated electronics, are summarized in Table~\ref{Table1}. 
\begin{table}[width=0.64\linewidth,cols=3,pos=h]
\centering
\caption{Technical specifications of the HPGe detector supplied by the manufacturer.}
\begin{tabular*}{\tblwidth}{@{}ll@{}}
\toprule  
Model                                  & GEM30P4-83-RB                 \\
Manufacturer                           & ORTEC                         \\
In service since                       & 2017                          \\
Capsule type                           & Pop top                       \\
HV Bias                                & +2800\,V                       \\
Crystal polarity                       & p-type                         \\
DAQ                                    & Analog                        \\
Shaping time                           & 6\,$\mu$s                     \\
Geometry                               & Closed end                    \\
Cryostat type                          & Vertical dipstick             \\
Cryostat diameter                      & 75\,mm                        \\
Crytal diameter                        & 62\,mm                        \\
\bottomrule 
\end{tabular*}
\label{Table1}
\end{table} 

The characterization measurements have been carried out using a set of standard sealed disk-type gamma sources of the active diameter of $\sim$6~mm and thickness of $\sim$1~mm with absolute strength of gamma sources $\leq$95~kBq within 5\% uncertainty. These sources cover a gamma ray energy range from 80.9~keV to 1408~keV, as shown in Table~\ref{Table2}.
\begin{table}[h!,width=0.5\linewidth,cols=4,pos=h]
\centering
\caption{Radioactive gamma-ray sources used in the measurements.}
\begin{tabular*}{\tblwidth}{@{}lllll@{}}
\toprule
Sources    & Gamma   & Decay         & Half-     & Branching         \\
           & Energy  & mode          & life      & Ratio             \\
           & (keV)   &               &           & (\%)              \\
\toprule
$^{109}$Cd & 88.0    & EC           & 461.4~d    & 3.64               \\ 
\midrule
$^{57}$Co  & 122.1   & $\beta^{-}$  & 271.7~d    & 85.60              \\ 
           & 136.5   & $\beta^{-}$  &            & 10.68              \\
\midrule
$^{133}$Ba & 80.9    & EC           & 10.5~y     & 32.90              \\ 
           & 276.4   &              &            & 7.160              \\
           & 302.9   &              &            & 18.34              \\          
           & 356.0   &              &            & 62.05              \\          
           & 383.9   &              &            & 8.940              \\           
\midrule
$^{22}$Na  & 511.0   & $\beta^{+}$  & 2.6~y      & 180.7              \\ 
           & 1274.5  &              &            & 99.94              \\  
\midrule
$^{137}$Cs & 661.7  & $\beta^{-}$  & 30.1~y      & 85.10               \\            
\midrule
$^{54}$Mn & 834.5   & EC           & 312.2~d     & 99.98               \\            
\midrule
$^{65}$Zn & 1115.5  & EC           & 243.9~d     & 50.04               \\           
\midrule
$^{60}$Co  & 1173.2  & $\beta^{-}$ & 1925.3~d    & 99.85              \\            
           & 1332.5  &             &             & 99.98              \\
\midrule
$^{152}$Eu & 121.7   & EC           & 13.5~y    & 28.53              \\                       
           & 244.7   & EC           &           & 7.550              \\                       
           & 344.4   & $\beta^{-}$  &           & 26.60              \\                       
           & 411.1   & $\beta^{-}$  &           & 2.240              \\                       
           & 444.0   & EC           &           & 0.298              \\                       
           & 779.0   & $\beta^{-}$  &           & 12.93              \\                       
           & 867.4   & EC           &           & 4.230              \\                       
           & 964.0   & EC           &           & 14.51              \\                       
           & 1085.8  & EC           &           & 10.11              \\                       
           & 1089.7  & $\beta^{-}$  &           & 1.734              \\                       
           & 1112.1  & EC           &           & 13.67              \\                       
           & 1213.0  & EC           &           & 1.415              \\                       
           & 1299.1  & $\beta^{-}$  &           & 1.633              \\                       
           & 1408.0  & EC           &           & 20.87              \\
\bottomrule 
\end{tabular*}
\label{Table2}
\end{table}
\begin{figure}[!ht]
\centering
\captionsetup{justification=justified,font=sf,labelfont=bf}
\includegraphics[width=0.70\linewidth,height=0.50\linewidth]{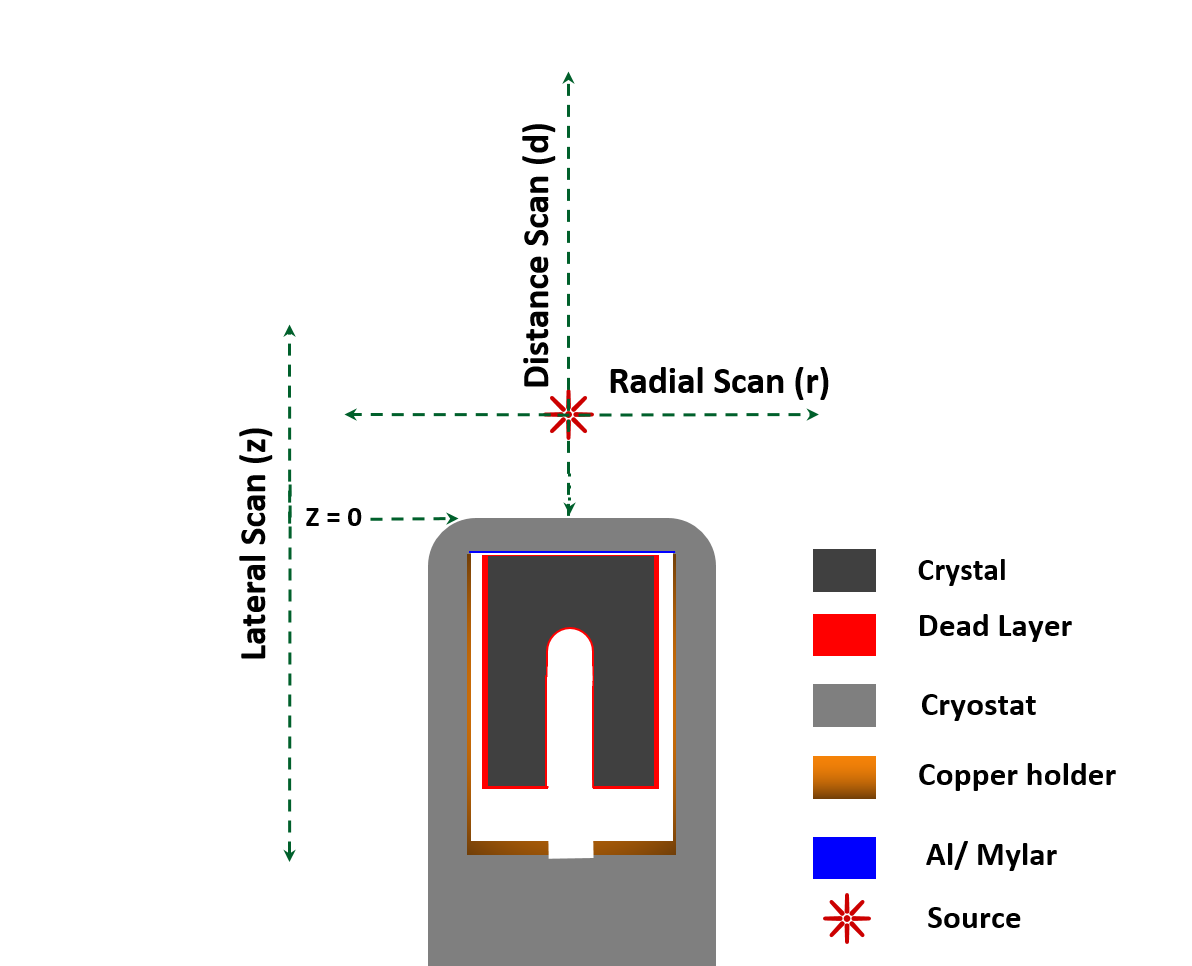}
\caption{A cross-sectional view of the detector with scanning directions indicated.}
\label{fig2}
\end{figure}
Radiography with X-rays and gamma-rays is used to determine the detector structure and its dimensions. However, an active volume may differ depending upon the electric field configuration inside the crystal \cite{boson2008,Pavel2008,Koleska2014}. In the present work, radiography is not possible, and hence mono-energetic gamma sources are used to scan the crystal in the energy range of 88-1332.5\,keV. In order to characterize the detector geometry, three types of scans, namely, i) Distance scan, ii) Radial scan, and iii) Lateral scan, have been performed around the detector. The scanning schemes of the detector are demonstrated in Figure~\ref{fig2}. The radial and lateral scans were carried out using $^{57}$Co, $^{60}$Co, $^{137}$Cs, $^{54}$Mn, $^{65}$Zn and $^{109}$Cd gamma sources, see Table~\ref{Table2} for the reference energies. The radial scan was performed by moving the source parallel to the top detector face at a distance of 10\,mm from the top face of the detector endcap (Carbon fiber window) in the position steps of 5\,mm. It covered a range of $\pm$\,6.5\,cm w.r.t. the center of the detector. For the lateral scan, the gamma source was moved parallel to its cylindrical axis at a distance of 10\,mm from the side face of the detector in 5\,mm steps and covered a range of -2\,cm to 9\,cm again w.r.t. the detector endcap. The distance scan was performed with point-like sources at a regular interval from 5\,cm to 30\,cm along the axial symmetry axis away from the top face of the detector. Typical uncertainty in the positioning of the source in all scans is estimated to be $\leq$1\,mm. Since some of the parameters are unknown or maybe not even related to the specific detector but are instead averaged over a set of detectors of the same or similar type, generic parameters offered by the manufacturer are typically insufficient. Therefore, parameters must be optimized. In order to determine the actual value of the parameters, optimizations are made to the crystal length (L), crystal radius (R), hole length ($L_{h}$), front gap, and dead layers.  
Given relatively low source strengths, no pileup effects have been observed in the spectra. Errors associated with the photopeak efficiency were computed, including statistical and systematic errors. Typical errors obtained in the present work were estimated to be $\sim$\,0.2\% in radial/lateral scans for $\textrm{E}_{\gamma}$ = 88\,keV, $\sim$\,0.5\% for $\textrm{E}_{\gamma}$ = 122.1\,keV, $\sim$\,0.4\% for $\textrm{E}_{\gamma}$ = 661.7\,keV and $\sim$\,0.1\% for $\textrm{E}_{\gamma}$ = 1115.5\,keV. It should be noted that the statistical errors are mainly due to the difference in strengths of various sources and energy-dependent variation in detection efficiency. Similarly, scan errors in the experiment were $\sim$\,0.1\% for the axial distances. Measurements were also performed with multi-gamma sources such as $^{60}$Co, $^{152}$Eu and $^{133}$Ba at z~$\geq$~10\,cm to ensure that the coincidence summing is negligible.
Data have been acquired for at least 10,000 counts for characterization measurements to reduce the statistical error to 1\%. Detector dead time has been monitored throughout the measurement and maintained to be less than 1\%. No significant worsening of energy resolution and shift in energy was observed at different times of long counting measurements of soil samples. The data of the background spectrum was recorded in a timestamp of 24~h. The raw data is then converted and analyzed using LAMPS~\cite{lamps} software. Each photopeak is fitted to the sum of Gaussian and second-order polynomials for a background to extract the net photopeak area. 

\begin{table}[h!,width=0.5\linewidth,cols=4,pos=h]
\centering
\caption{Typical clay loam composition of of soil samples ~\cite{Medhat2014}.}
\begin{tabular*}{\tblwidth}{@{}cccc@{}}
\toprule
Compound         & Mass fraction   & Compound    & Mass fraction \\
\toprule
SiO$_{2}$        & 0.5890          & K$_{2}$O    & 0.0325  \\ 
\midrule
Al$_{2}$O$_{3}$  & 0.1625          & Na$_{2}$O   & 0.0235 \\ 
\midrule
Fe$_{2}$O$_{3}$  & 0.1340          & MgO         & 0.0135 \\
\midrule
CaO              & 0.0360          & TiO$_{2}$   & 0.0090 \\
\bottomrule 
\end{tabular*}
\label{Table3}
\end{table}
\section {Results and Analysis} \label{Results and Analysis}

\subsection{Spectroscopic performance of the detector}
The accuracy of the measurements depends profoundly on the performance and stability of the detector and associated electronics. Different characteristics of the HPGe detector, such as energy calibration and resolution, peak shape and Peak-to-Compton ratio (P/C), Full energy peak efficiency (FEPE), and Figure of Merit (FoM), are determined as a function of gamma-ray energies against the warranted values provided by the manufacturer. The radioactive gamma sources used in the scanning of the detector are given in Table~\ref{Table2}. For the present HPGe detector, R.E. was found to be 33 (0.3)\% in the laboratory test, which is 6\% lower than the measured value provided by the manufacturer.
The HPGe detector has been calibrated using different standard gamma sources and shown in Figure~\ref{fig3}~(a). The line through the data points represents a fitting function of type y = a + {${\textrm{b}} {{x}}$, where a = 0.4 $\pm$ 0.03 and b = 0.38185 $\pm$ 1.59279 $\times 10^{-5}$. As can be noticed from this figure, a good linear relationship with the channel number can be observed. The statistical correlation coefficient is found to be 1 for each measured data point with the detector. The energy resolution measures the width (FWHM) of a single energy peak at a specific energy, usually expressed in~keV for germanium detectors. It may be pointed out that the typical energy resolution of NaI and HPGe detectors are found to be 50\,keV and 1.5\,keV at 1332.5\,keV, respectively~\cite{Gilmore}. Hence, HPGe is preferred over NaI detectors for high-resolution gamma-ray spectroscopy, even though the NaI is likely to have greater counting efficiency. Generally, the energy resolution of a detector is expressed as the ratio of FWHM to the gamma-ray energy. The distribution of energy resolution with different gamma-rays is shown in Figure~\ref{fig3}~(b) and fitted to an empirical three-parameter function of type, 
\begin{equation}\label{c2:eqn:1}  
\textrm{R} = \frac{{\textrm{A}}}{{\textrm{E}}^{{\textrm{B}}}} + \textrm{C}
\end{equation}

The best-fit values to the fitted parameters A, B, and C are 0.4 $\pm$ 0.03, 0.9 $\pm$ 0.01 and 6.42229 $\times 10^{-4}$ $\pm$ 3.43594 $\times 10^{-5}$, where A and B are in~keV. The typical energy resolution of this detector at 1332.5\,keV is 1.72~keV and found to be similar to another LN$_2$ based HPGe detector of similar relative efficiency present in the laboratory. No worsening of the energy resolution has been observed over the running period of about five years. In addition to the FWHM taken at each peak, the full width at one-fifth maximum (FW.2M), full width at one-tenth maximum (FW.1M), and full width at one-fiftieth maximum (FW.02M) are also recorded to check for the worsening of the tail.
\begin{figure}[h!]
\centering
\captionsetup{justification=centering}
\begin{tabular}{@{}c@{}}
\includegraphics[width=8.5cm,height=5.6cm]{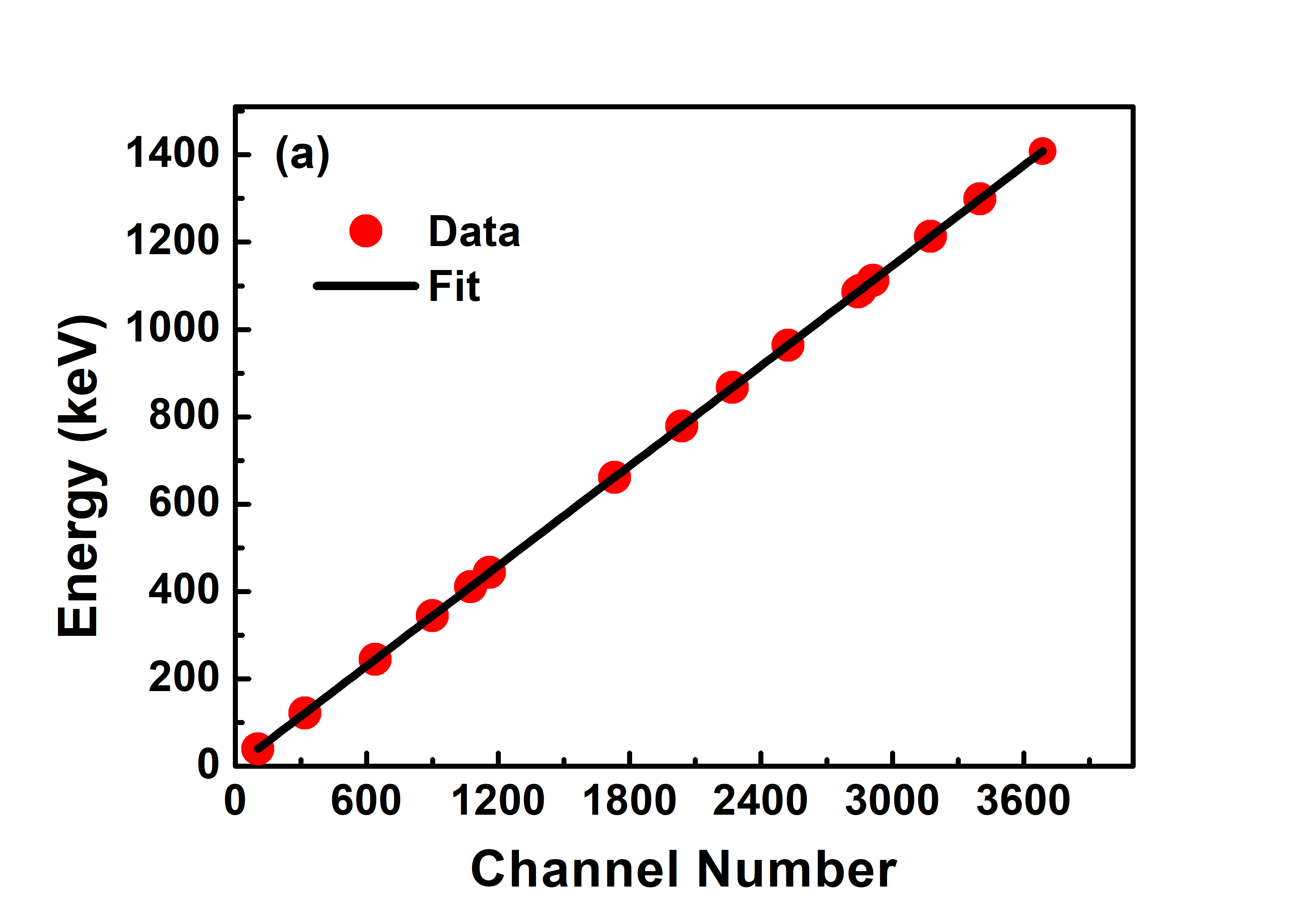}
\includegraphics[width=8.5cm,height=5.6cm]{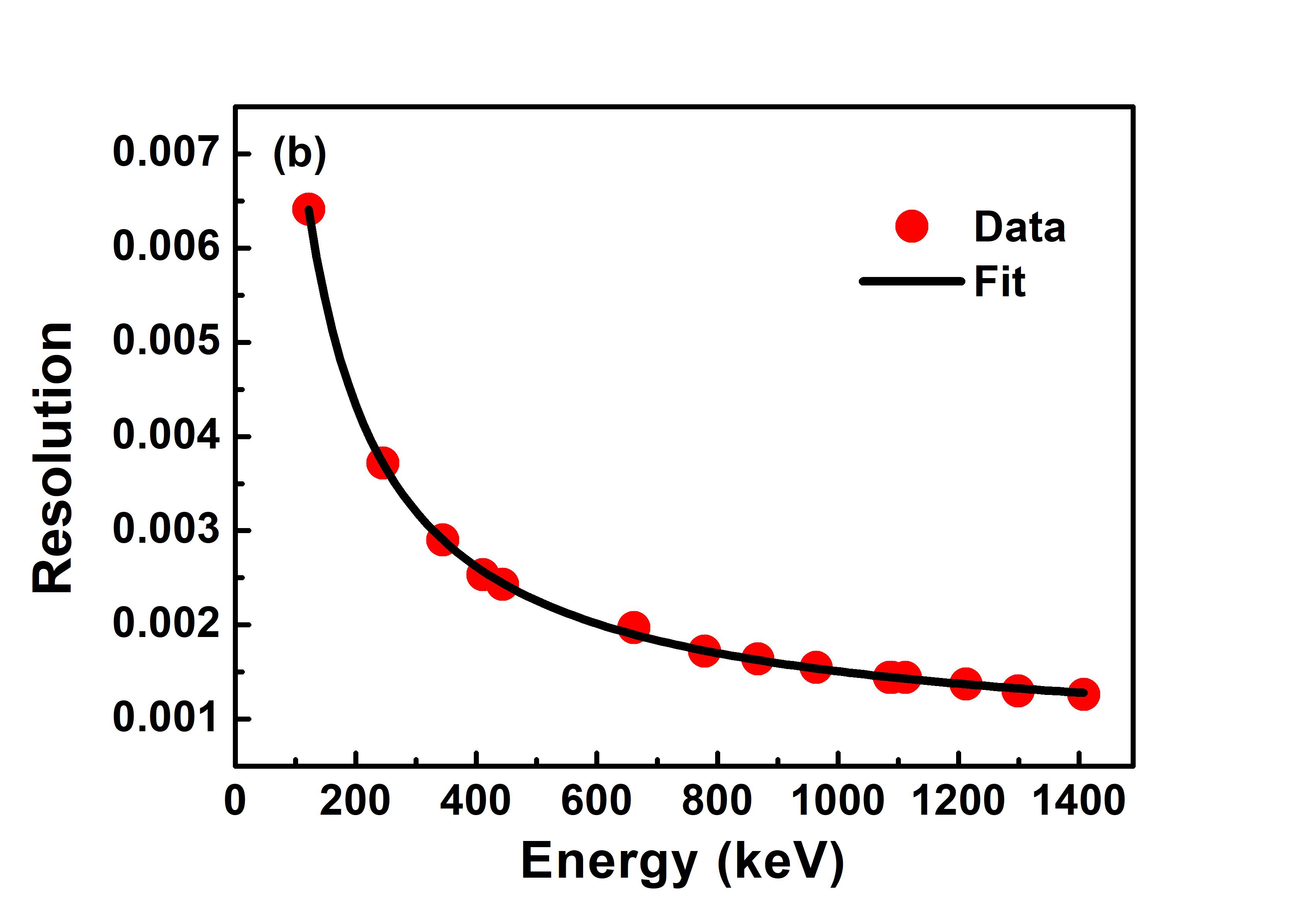}
\end{tabular}
\caption{Energy calibration and resolution of the HPGe detector are shown in (a) and (b), respectively. The solid line through the data points is the best fit.} 
\label{fig3}
\end{figure}

The peak-to-Compton ratio has been measured from the same gamma line, 1332.5\,keV, which has been used for the resolution measurement obtained from $^{60}$Co source. The value of this ratio was measured at various axial distances, and the mean value of the peak-to-Compton ratio was found to be 62:1, following the manufacturer data.

The efficiency calibration of a spectrometer is of great importance in analyzing radionuclides of interest. Different physical parameters, such as; the crystal volume and shape, source dimensions, gamma-ray absorption cross-section, attenuation layers in front of the detector, and the distance and position from the source to the detector, determine the efficiency of the detector~\cite{boson2008,Pavel2008,2009budjas}. Absolute efficiency, also known as full energy peak efficiency (FEPE), is defined as the ratio of the number of photopeak counts detected to the total number emitted by the source and can be determined according to the following equation:
\begin{equation}\label{c2:eqn:2}
\varepsilon =\frac{\textrm{N$_{\gamma}$}}{\textrm{A$_{\gamma}$} \ast \textrm{I$_{\gamma}$} \ast \textrm{t}}
\end{equation}
Where N$_{\gamma}$ is the net photopeak area (background subtracted), A$_{\gamma}$ is the present activity of source (Bq), I$_{\gamma}$ is the gamma-ray emission probability. t is the time elapsed (taking into account the analyzer counting losses). Figure~\ref{fig4} shows the efficiency measurements for all the considered gamma-ray energies at a distance of 25~cm.
\begin{figure}[!ht]
\centering
\captionsetup{justification=justified,font=sf,labelfont=bf}
\includegraphics[width=0.58\linewidth,height=0.40\linewidth]{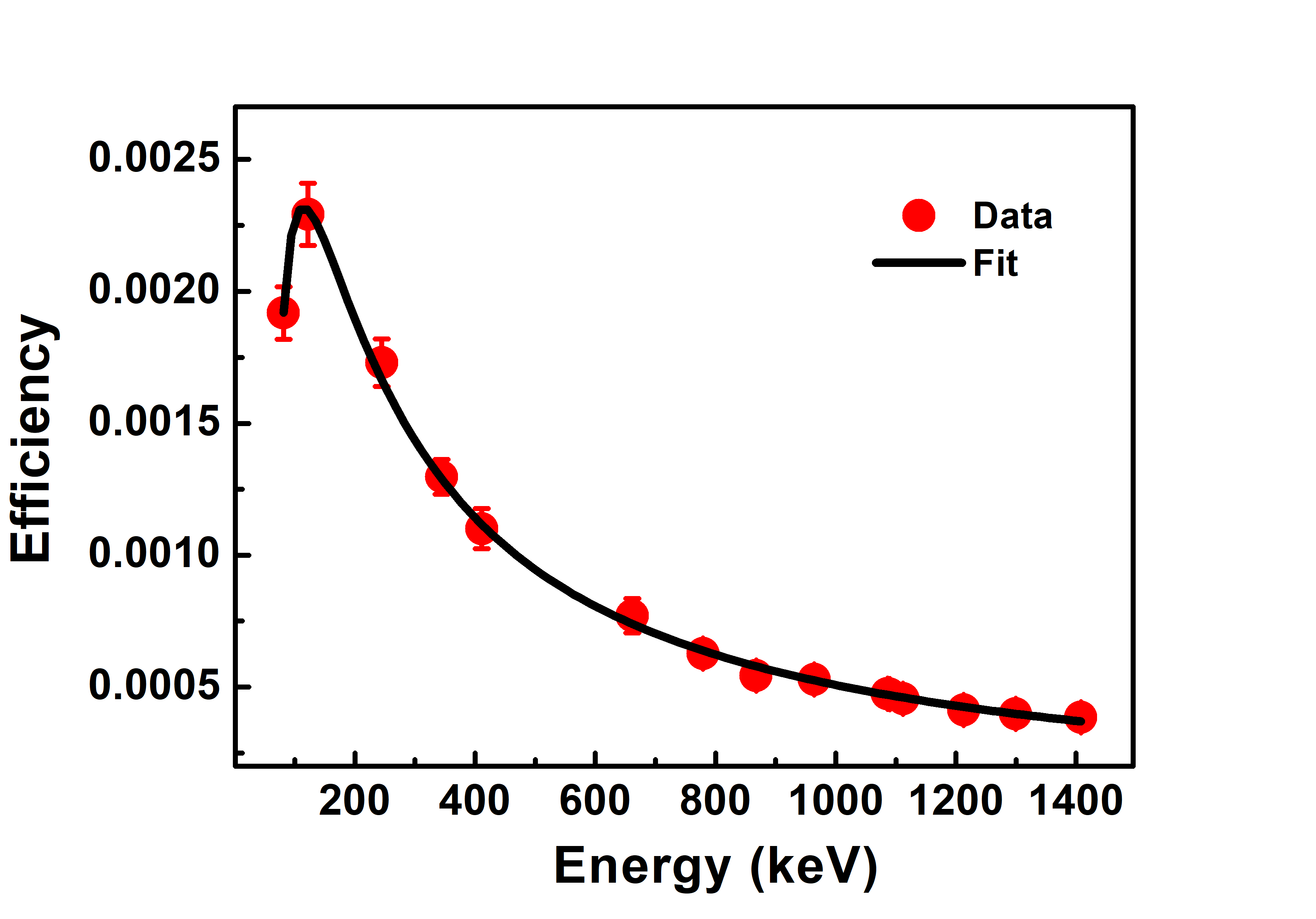}
\caption{Efficiency calibration curve of the HPGe detector at source to detector distance of 25~cm.}
\label{fig4}
\end{figure}
The quoted errors may be due to the number of counts in the photopeak (statistical) and source activities (systematic). As shown in Figure~\ref{fig4}, the solid line through the data points represents the best fit to the data. The efficiency curve is fitted using an inverse square fitting function of type,
\begin{equation}\label{c2:eqn:3}  
\textrm{$\varepsilon$} =\textrm{A} + \frac{\textrm{B}}{\textrm{E}} + \frac{\textrm{C}}{\textrm{E}^{{2}}} 
\end{equation}
where $\varepsilon$ is the photopeak efficiency, and E is the gamma-ray energy. The best-fit values to the fitted parameters A, B, and C are found to be 1.23209 $\times 10^{-5}$ $\pm$ 0.91263 $\times 10^{-5}$, 0.5 $\pm$ 0.01 and -30 $\pm$ 1. The fitting function yields good approximations over different energy ranges and for different crystal sizes. As can be noticed from the figure, some fluctuations in the data points are primarily due to variations in peak shape and low count rates. The HPGe detector has a closed-end coaxial configuration of p-type material and a thick attenuation layer at the detector entrance contact. Correspondingly, the detector's efficiency was found to be less for low energy gamma-rays below 80.9\,keV as clearly visible in Figure~\ref{fig4}. The efficiency curve shows a rapid increase from 80.9\,keV of $^{133}$Ba source, peaks at 121.8\,keV of $^{152}$Eu, and sharply decreases as a result of an increase in gamma-ray energy. This implies that the efficiency is maximum at low energy and decreases exponentially at higher emitted gamma-rays consistent with similar detectors by ORTEC~\cite{Aguilar2020}. The efficiency curve demonstrates the excellent performance of the measurements and analysis applied in this work. Measurement of photopeak efficiencies using the above radioactive sources can help better estimate the detector's active volume and surrounding materials. 

\subsection{Detector modelling}\label{Detector modelling}
The Monte Carlo simulation program was developed using the GEANT4 framework (version 4.10.00.p02). The photopeak efficiencies have been estimated directly by determining the energy deposition in the simulated volume of the detector. The source-detector geometry, as shown in Figure~\ref{fig2}, has been implemented using the information provided by the manufacturer. The primary particles were generated from the input files using G4ParticleGenerator class. The 10$^6$ photons were generated uniformly from the radioactive source to keep the statistical error below 1\%. The photons cutoff energy was set at 1~keV using the electromagnetic processes PhysicsList. The energy calibration obtained from the experiments has been utilized to set up the energy bins in the simulated spectra. In order to avoid summing effects, the photopeak efficiencies were calculated using the mono-energetic sources at close distances. The photons that entirely lose their energy in the active volume of the detector are taken into account for evaluating photopeak efficiencies. The efficiency at the given energy was calculated by generating the histogram for a number of events against the energy deposited inside the active volume of the crystal. The simulation results have been analyzed using the ROOT data analysis framework. The simulated photopeak efficiency ({\textrm{$\varepsilon$$^{\textrm{Sim}}$}}) has been determined from the stored energy histograms using the following equation,
\begin{equation}\label{c2:eqn:4}
\textrm{$\varepsilon$$^{\textrm{Sim}}$} = \frac{\textrm{N$_{c}$}}{\textrm{N$_{g}$} }
\end{equation}
Where {\textrm{N$_{c}$}} is the number of photons that deposited energy in the crystal after the background elimination and {\textrm{N$_{g}$}} is the number of photons generated.
The relative deviations (\textrm{$\sigma$$_{R}$}) between simulated and experimental photopeak efficiencies were calculated using equation{~\ref{c2:eqn:5}, where n is the total number of data points in each set corresponding to a specific energy and {$\textrm{r}_i$} represents the individual data points in the data set.
\begin{equation}\label{c2:eqn:5}
\textrm{$\sigma$$_{\textrm{R}}$} = \frac{1}{\textrm{n}}{\Sigma_{i=1}^\textrm{n} {\frac{{\textrm{$\varepsilon$$^{\textrm{Exp}}$}  (\textrm{$\textrm{r}_i$}}) - {\textrm{$\varepsilon$$^{\textrm{Sim}}$}  (\textrm{r}_i)}}{\textrm{$\varepsilon$$^{\textrm{Sim}}$}  (\textrm{r}_i)}}}
\end{equation}
Equation~\ref{c2:eqn:6} defines the total relative deviation (\textrm{$\sigma$$_{\textrm{TR}}$}), where m is the number of data sets that correspond to the various energy scans and $\textrm{$\sigma$$_{R}$} (\textrm{E}_i)$ is the relative deviation corresponding to the different energies.
\begin{equation}\label{c2:eqn:6}
\textrm{$\sigma$$_{\textrm{TR}}$} = \frac{1}{\textrm{m}}{\Sigma_{i=1}^\textrm{m} \textrm{$\sigma$$_{\textrm{R}}$} (\textrm{E}_i)}
\end{equation}
\begin{figure}[h!]
\centering
\captionsetup{justification=centering}
\begin{tabular}{@{}c@{}}
\includegraphics[width=8.5cm,height=5.6cm]{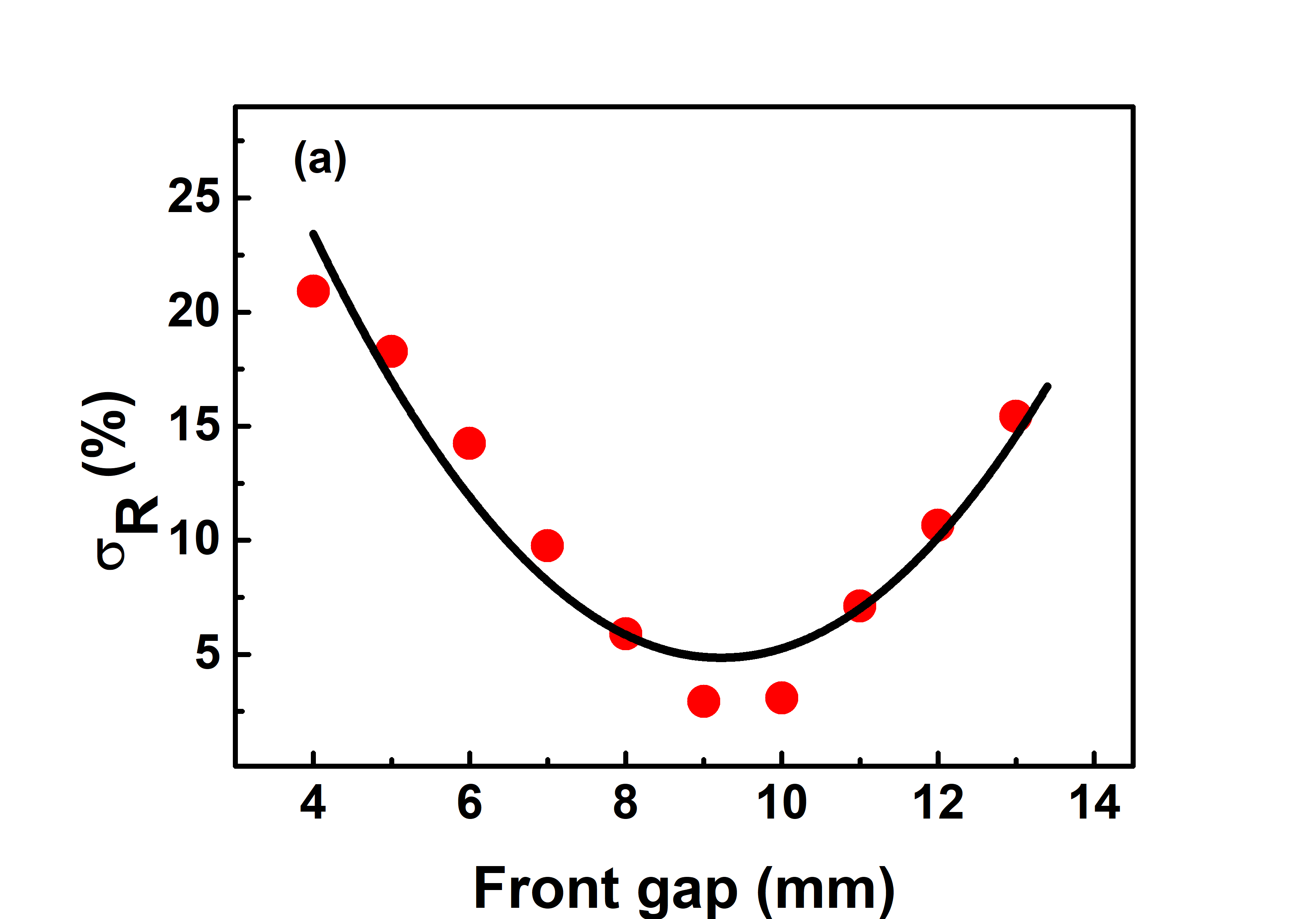}
\includegraphics[width=8.5cm,height=5.6cm]{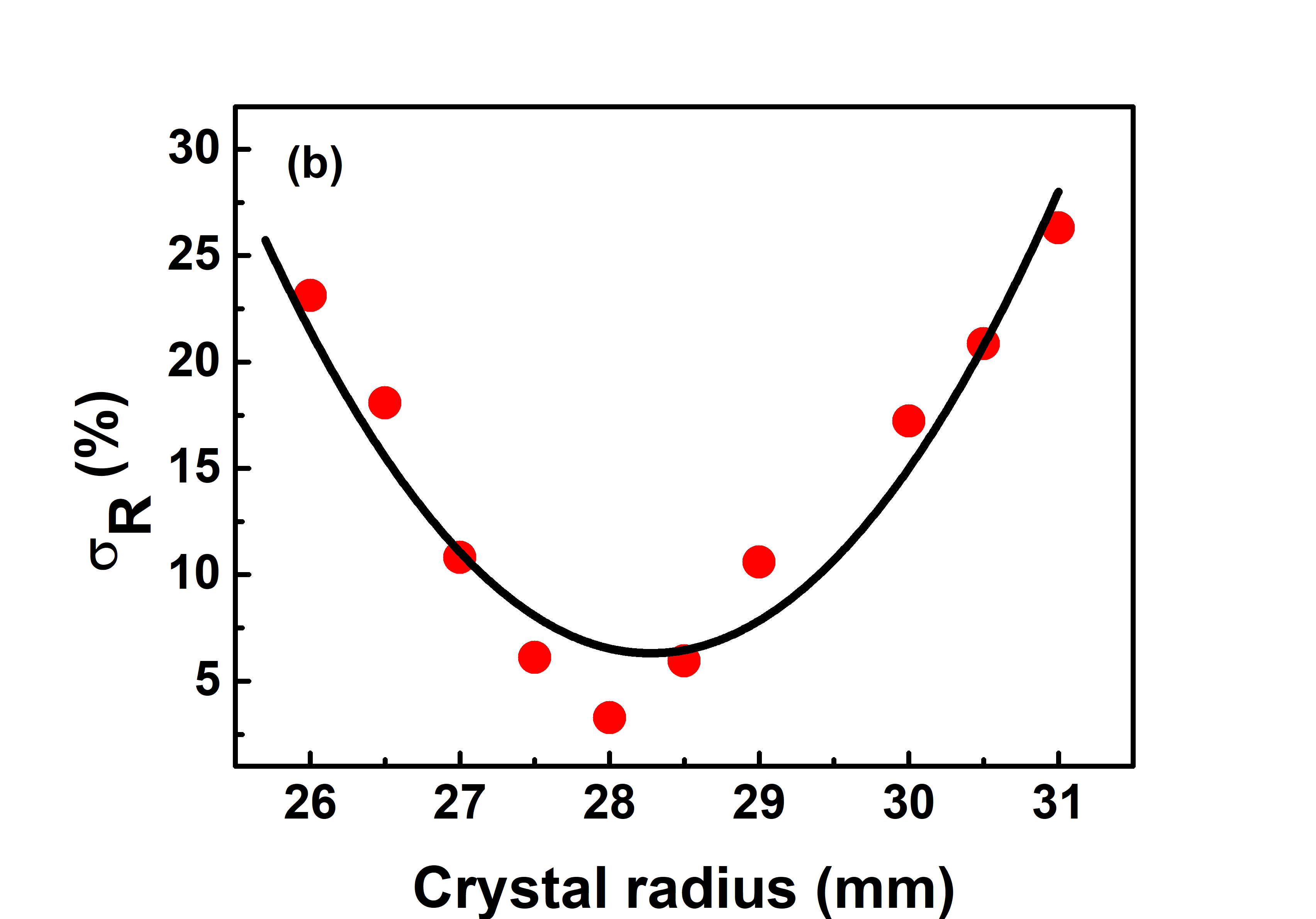}
\end{tabular}
\caption{Relative deviation (\textrm{$\sigma$$_{R}$}) as a function of (a) front gap (g) and (b) crystal radius (R) for a radial scan with 1115.5~keV for the initial estimate of g and R.}
\label{fig5}
\end{figure}
A polynomial function is used to fit the given data points and determine the minimum value of \textrm{$\sigma$$_{\textrm{R}}$} that yields the optimal fit. As low-order polynomials exhibit greater smoothness in comparison to high-order polynomials, the parabolic \textrm{{$\sigma$$_\textrm{R}$} = {${\textrm{a}}{{x}^2}$} + {${\textrm{b}} {{x}}$} + c } function was used to fit the given data. As an illustration, the best-fit values of front gap (g) and crystal radius (R) are shown in Figure~\ref{fig5}~(a) and Figure~\ref{fig5}~(b), initial estimation was done using $^{65}$Zn source.

\subsubsection{Parametric study} \label{Parametric study}
Initially, the detector model was constructed according to the specifications provided by the manufacturer. Step-by-step adjustments were carried out to optimize the parameters, namely, crystal radius (R), hole depth (L$_{h}$), dead layers (t$_{d}$, t$_{b}$, and t$_{s}$) and front gap (g) to obtain an effective detector model (see Table~\ref{Table4}). 
\begin{table}[width=0.75\linewidth,cols=4,pos=h]
\centering
\caption{List of detector parameters, range, and step size used in the simulations. Parameters marked in $\oplus$ are unaltered in the present work.}
\begin{tabular*}{\tblwidth}{@{}lllll@{}}
\toprule  
Parameter                       & Manufacturer      & Varied from          & Step size      & Optimized     \\
                                & (mm)              & (mm)                 & (mm)           & (mm)         \\
\toprule                                    
Crystal Radius (R)              &  31               &  25-31               & 0.5             & 29.53 $\pm$ 1.2 \\
Hole radius$^{\oplus}$ (h$_r$)  &  5.55             &    --                &   --            & 5.55 \\
Disc length (L$_1$)             &  6.12             &  5.3-6.1             &  0.1            & 5.83 $\pm$ 0.1\\
Hole length (L$_h$)             &  33.7             &  30.4-34.4           &  0.4            & 32.4\\
Crystal length (L)              &  46               &  42-46               &  1              & 44 $\pm$ 0.3 \\
Top dead layer (t$_d$)          &  --               &  0-2.6               &  0.3            & 1.8 $\pm$ 0.2 \\
Side dead layer (t$_s$)$^{\ast}$&  0.7              &  --                  &  --             &1.7\\
Bottom dead layer (t$_b$)       &  --               &  --                  &  --             & 2 \\
Front gap (g)                   &  4                &  4-13                & 1              & 9 $\pm$ 1.0 \\
Front carbon fiber$^{\oplus}$   &  0.9              &   --                 &    --           &0.9 \\
Side carbon fiber$^{\oplus}$    &  1.6              &   --                 &    --           &1.6 \\
Cu cup thickness$^{\oplus}$     &  0.8              &   --                 &    --           &0.8 \\
Ge crystal volume (V)           &  135 \textrm{$\textrm{cm}^3$}      &   --                 &    --           & 120 $\pm$ 4.8  $\textrm{cm}^3$    \\
\bottomrule
{\footnotesize $^{\ast}$ similar to CRADLE~\cite{gupta2018}}
\end{tabular*}
\label{Table4}
\end{table}
To replicate the experimental data, the detector's size must be optimized. The two main dimensions that have a significant impact on volume are R, and L. Crystal length (L) was constructed as solid and hollow cylinders and optimized in two parts, namely, Disc length (L$_{1}$) and hole depth (L$_{h}$) and optimized independently. Low-energy gamma-rays are considered optimal for estimating the thickness of the top dead layer (t$_{d}$) on a crystal, as this layer attenuates gamma-rays. This has negligible effects on photopeak efficiencies at higher energies. Therefore, 88 and 122.1~keV energies are used for top dead layer determination. Depending on how long the detector has been in operation, it can decrease its active volume. The manufacturer did not supply the t$_{d}$ value. Because both detectors are identical, the same side dead layer (t$_{s}$) as CRADLE~\cite{gupta2018}} was implemented. A passive Ge dead layer with a thickness of b=L-{L$_{opt}$} is introduced into the model when employing the manufacturer's specified physical length, L. In the simulation, uniform dead layers were employed. The effect of air gap has significant effects on small source-detector distances. However, influence can be considered negligible at large distances. It can also be estimated better with low-energy gamma-rays. The results of the lateral scan, as shown in Figure~\ref{fig7}, indicate a significant deviation at the end side of the crystal. However, this issue was successfully addressed after the integration of the bottom dead layer into the model. 
\begin{figure}[h!]
\centering
\captionsetup{justification=centering}
\begin{tabular}{@{}c@{}}
\includegraphics[width=8.5cm,height=5.8cm]{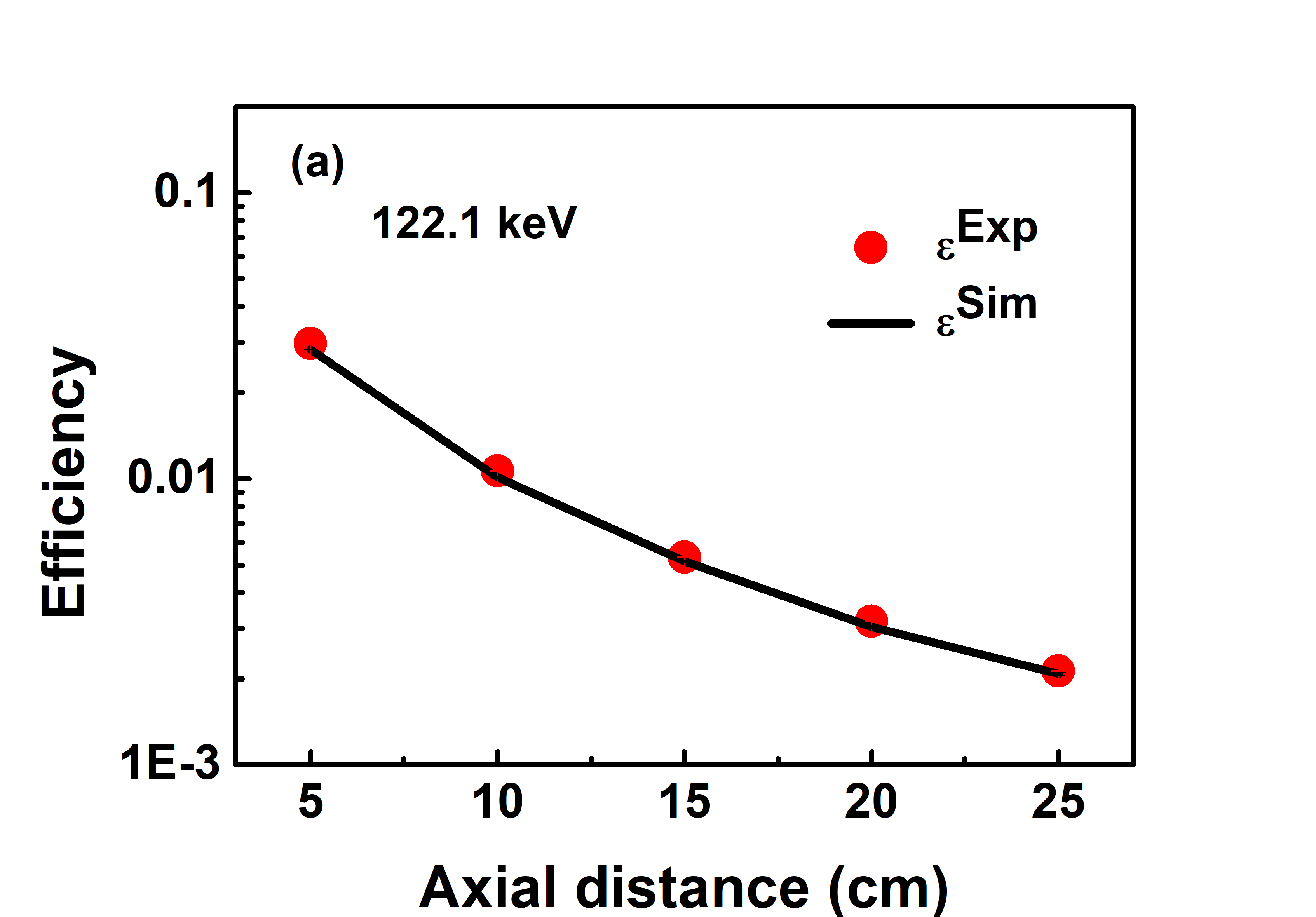}
\includegraphics[width=8.5cm,height=5.8cm]{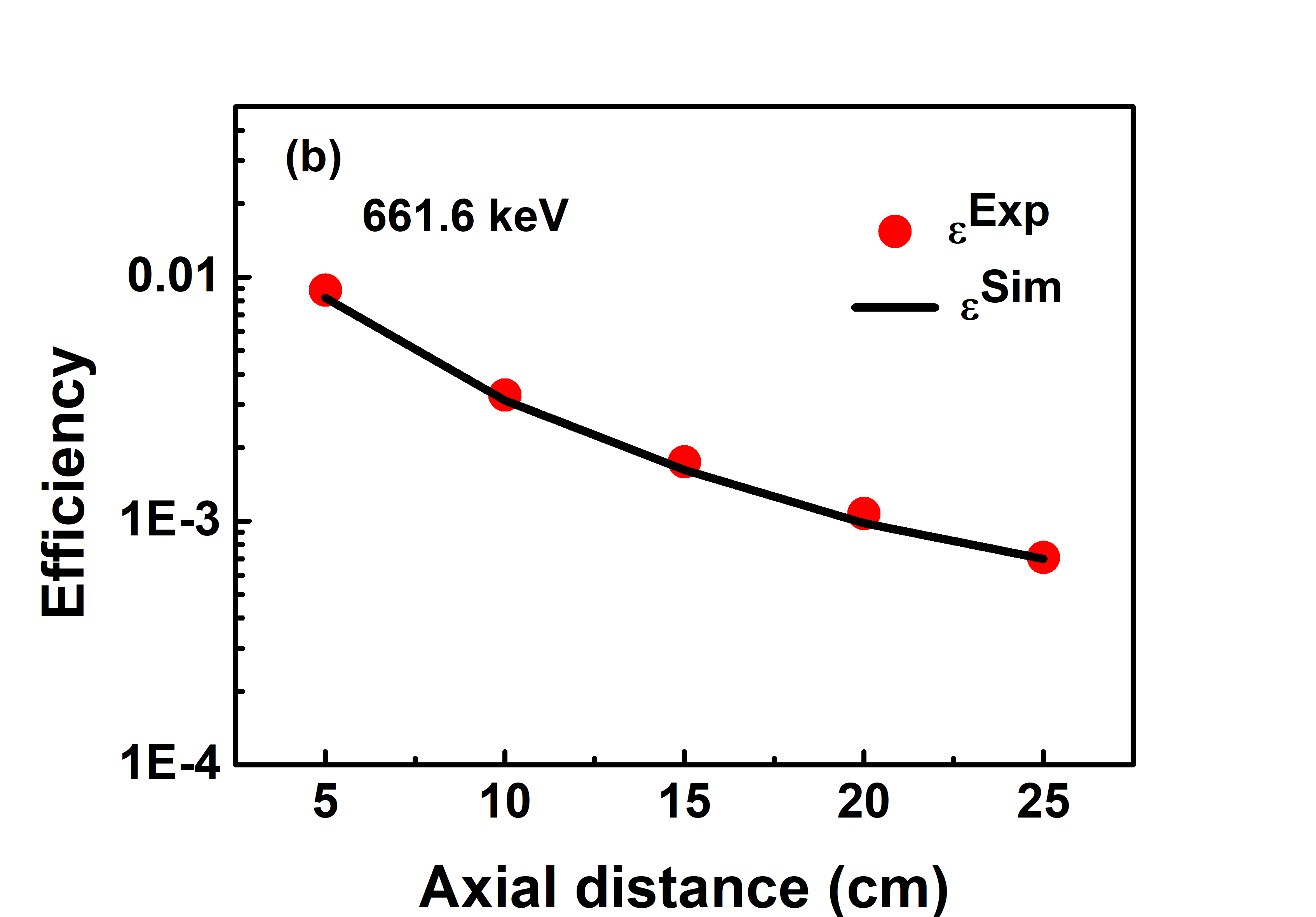}\\
\includegraphics[width=8.5cm,height=5.8cm]{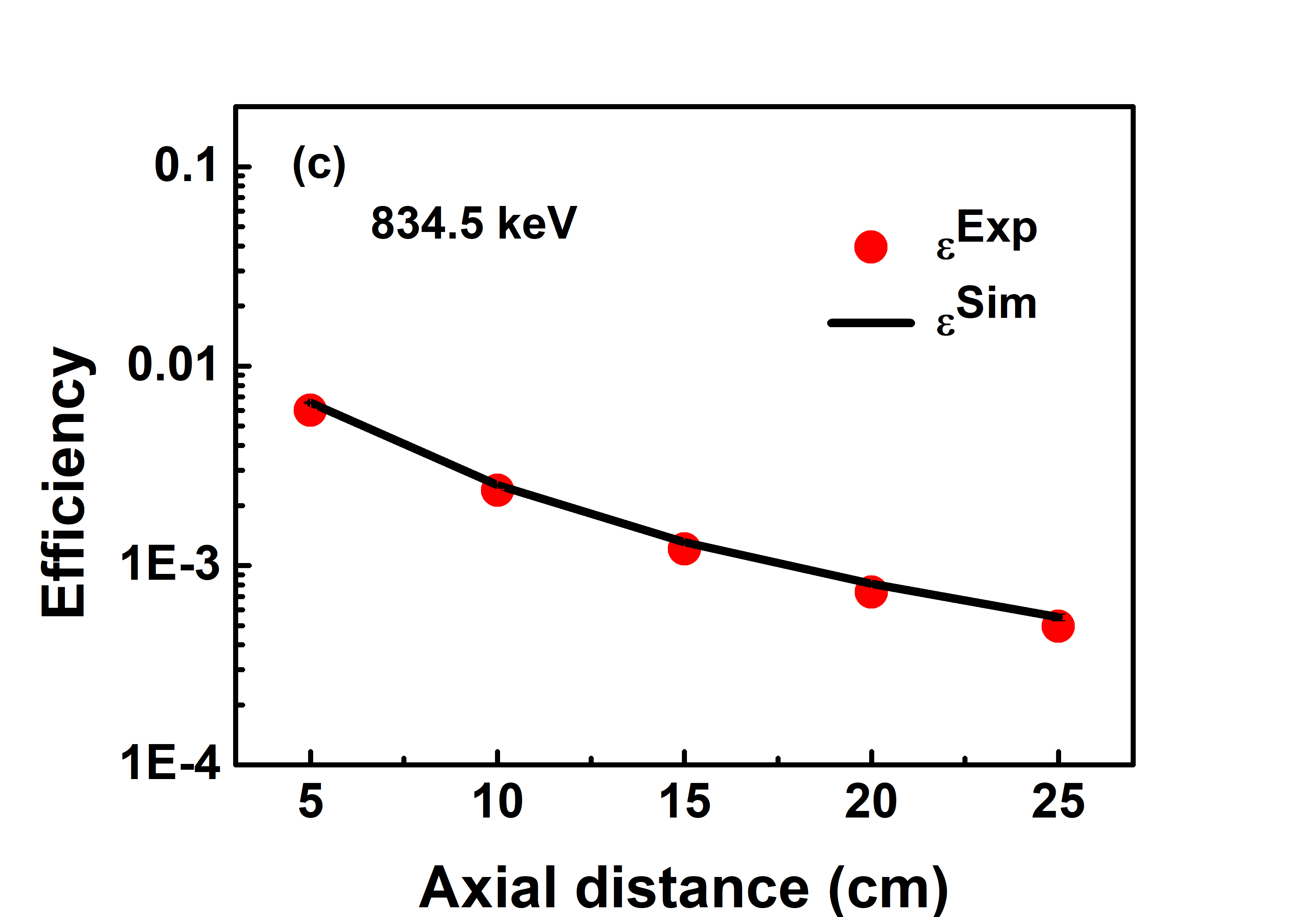}
\includegraphics[width=8.5cm,height=5.8cm]{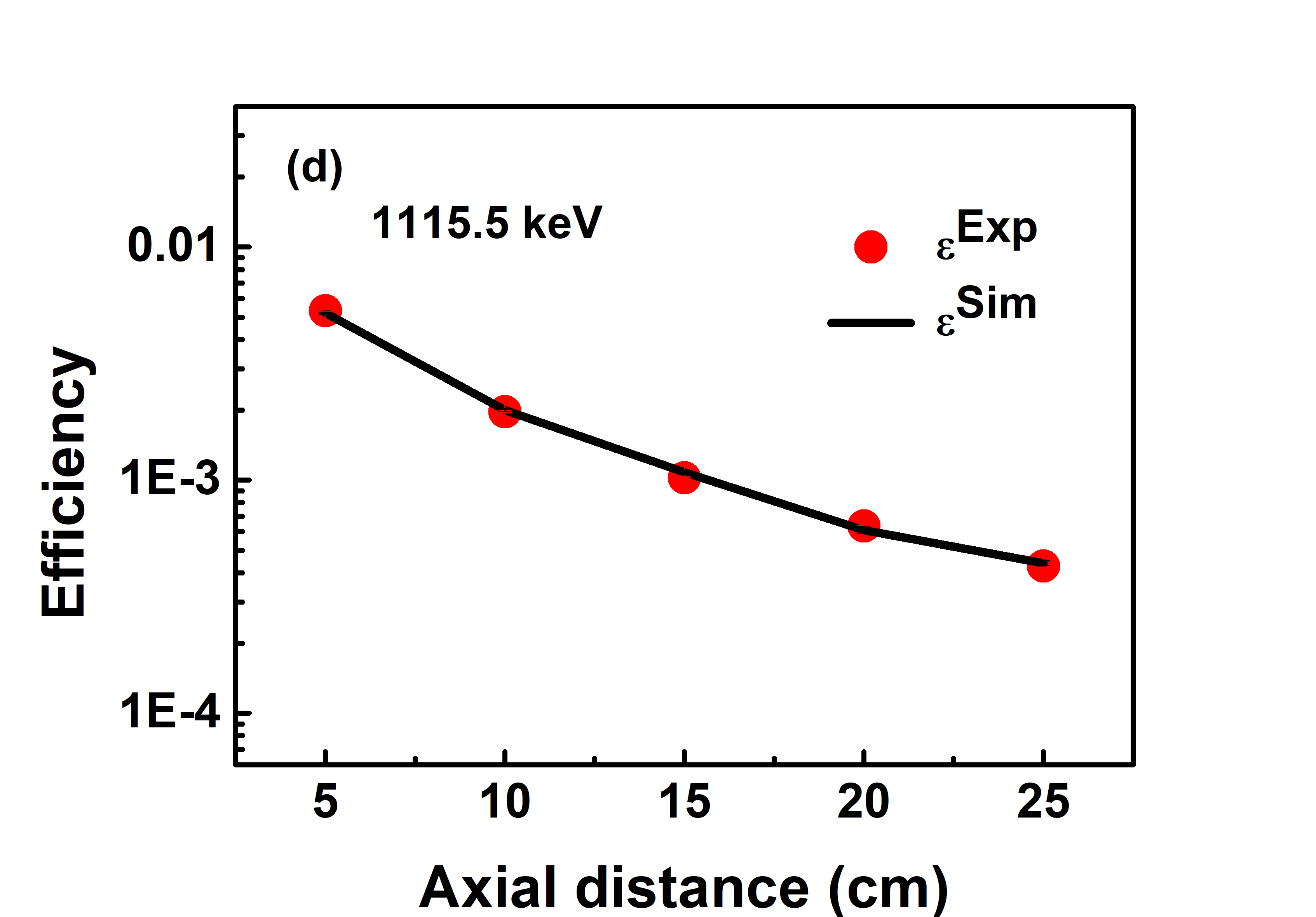}\\
\end{tabular}
\caption{Simulated and experimental absolute photopeak efficiency as a function of distance scan for 122.1~keV, 661.6~keV, 834.5~keV and 1115.5~keV (a), (b), (c) and (d) respectively.} 
\label{fig6}
\end{figure}
\begin{figure}[h!]
\centering
\captionsetup{justification=centering}
\begin{tabular}{@{}c@{}}
\includegraphics[width=8.5cm,height=5.8cm]{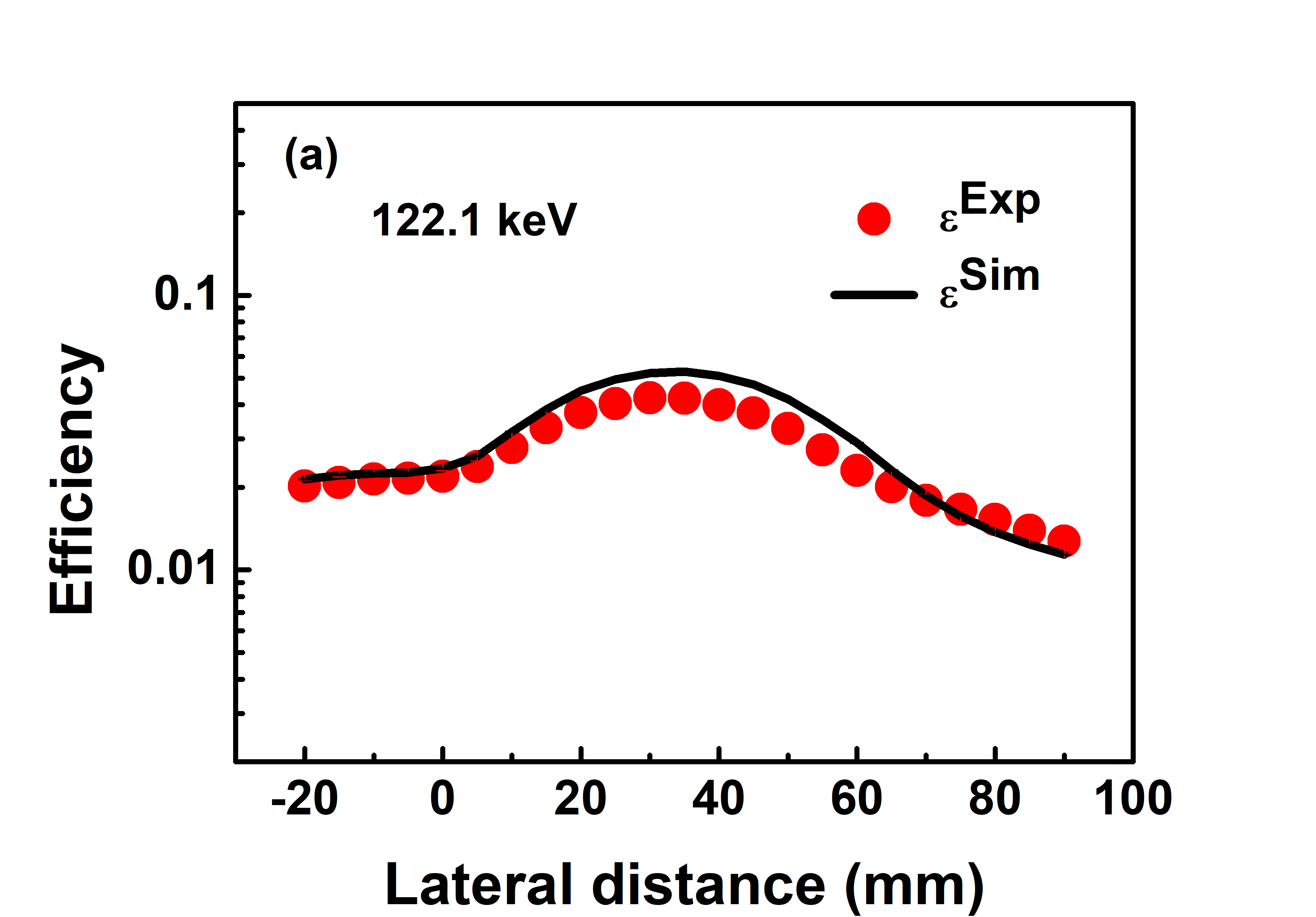}
\includegraphics[width=8.5cm,height=5.8cm]{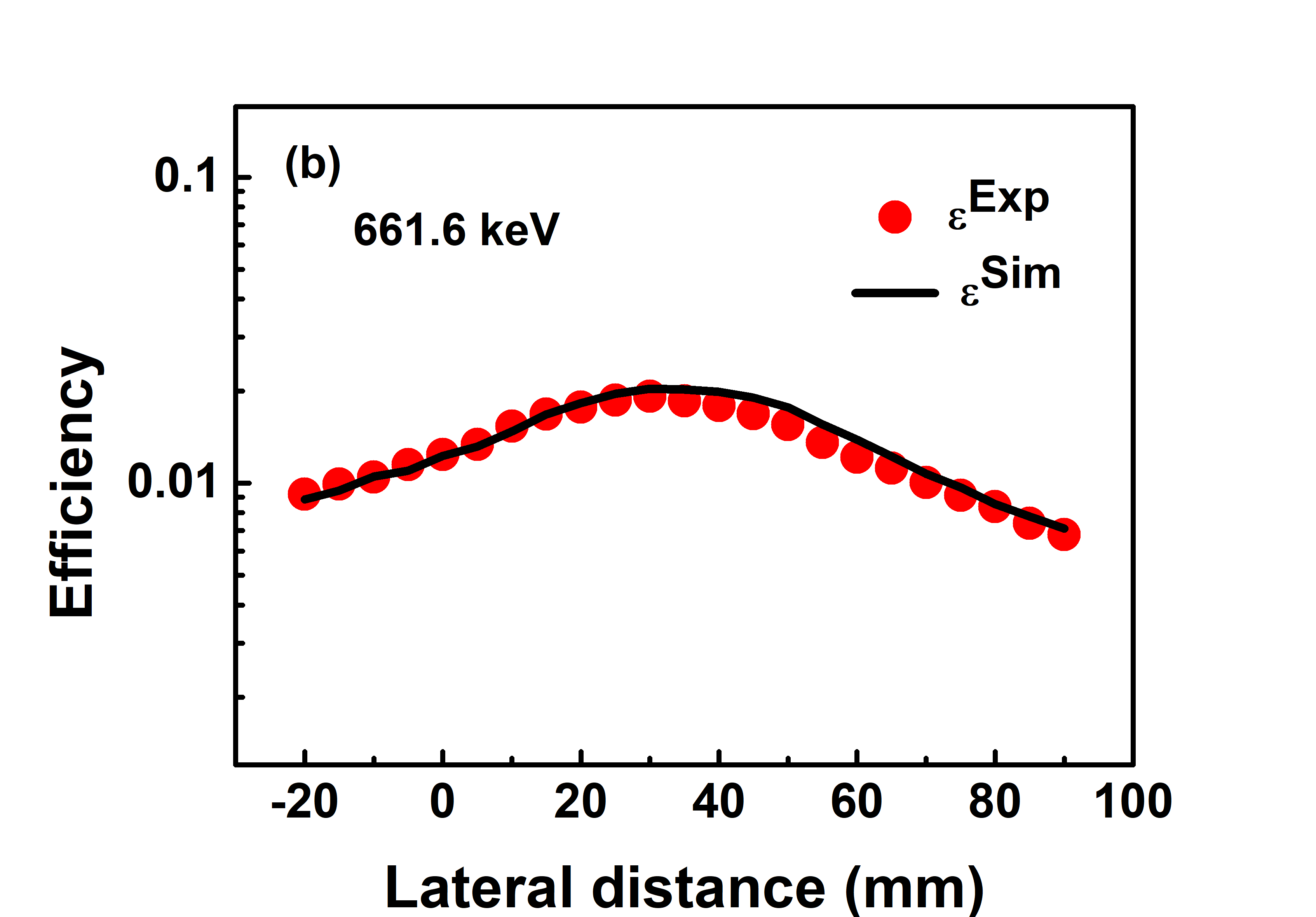}\\
\includegraphics[width=8.5cm,height=5.8cm]{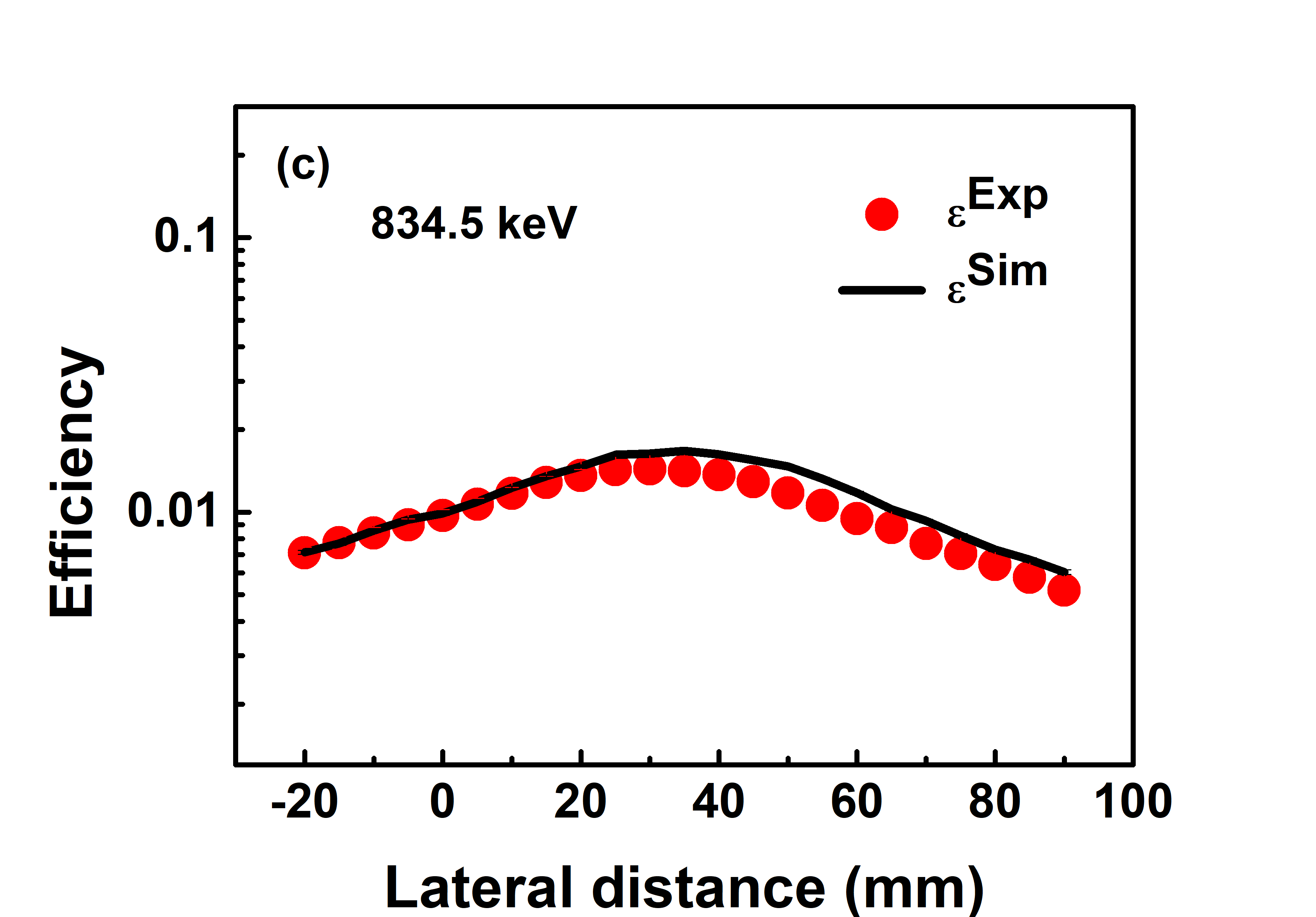}
\includegraphics[width=8.5cm,height=5.8cm]{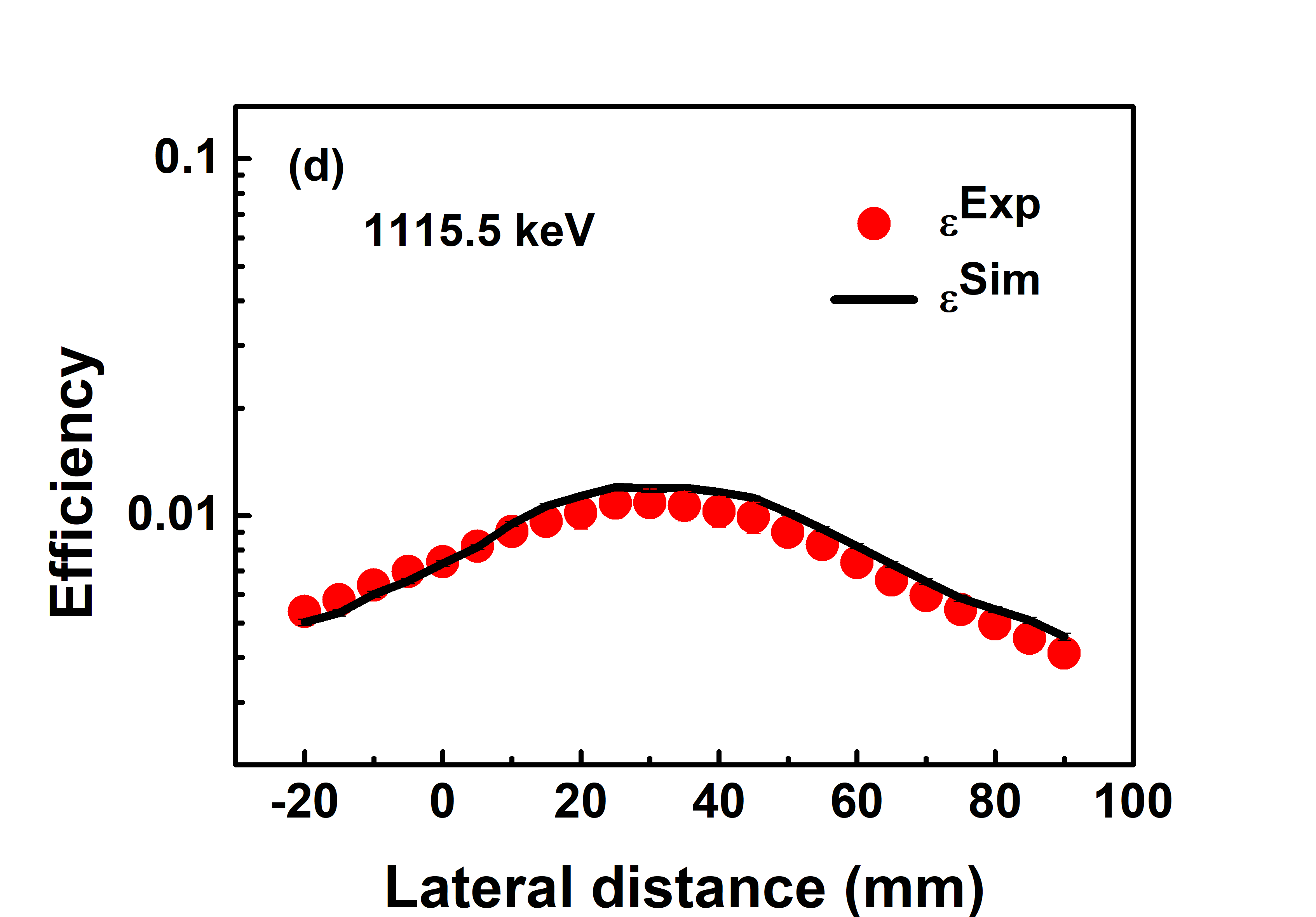}\\
\end{tabular}
\caption{Simulated and experimental absolute photopeak efficiency as a function of lateral scan for 122.1~keV, 661.6~keV, 834.5~keV and 1115.5~keV (a),(b), (c) and (d) respectively.}
\label{fig7}
\end{figure}
\begin{figure}[h!]
\centering
\captionsetup{justification=centering}
\begin{tabular}{@{}c@{}}
\includegraphics[width=8.5cm,height=5.8cm]{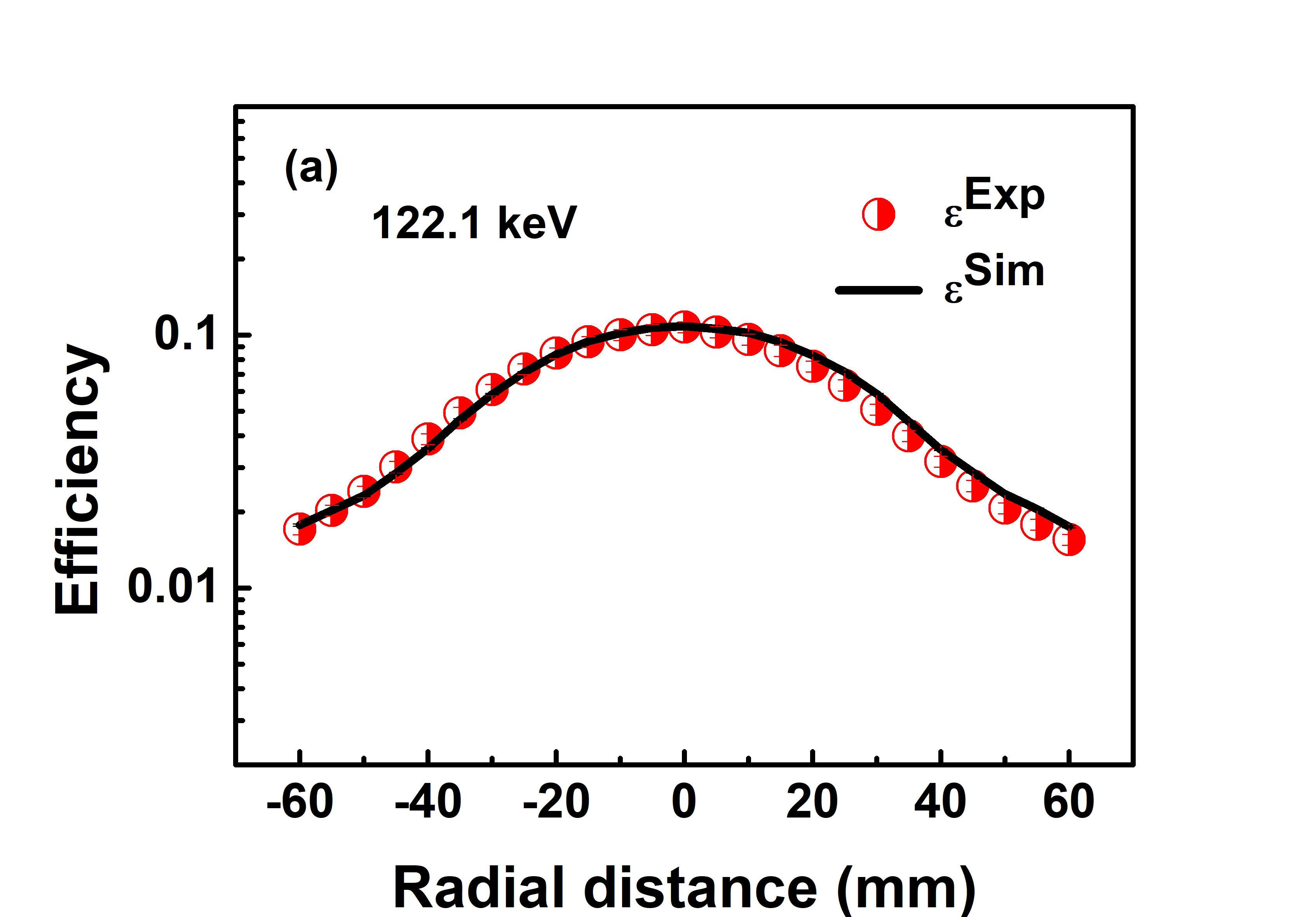}
\includegraphics[width=8.5cm,height=5.8cm]{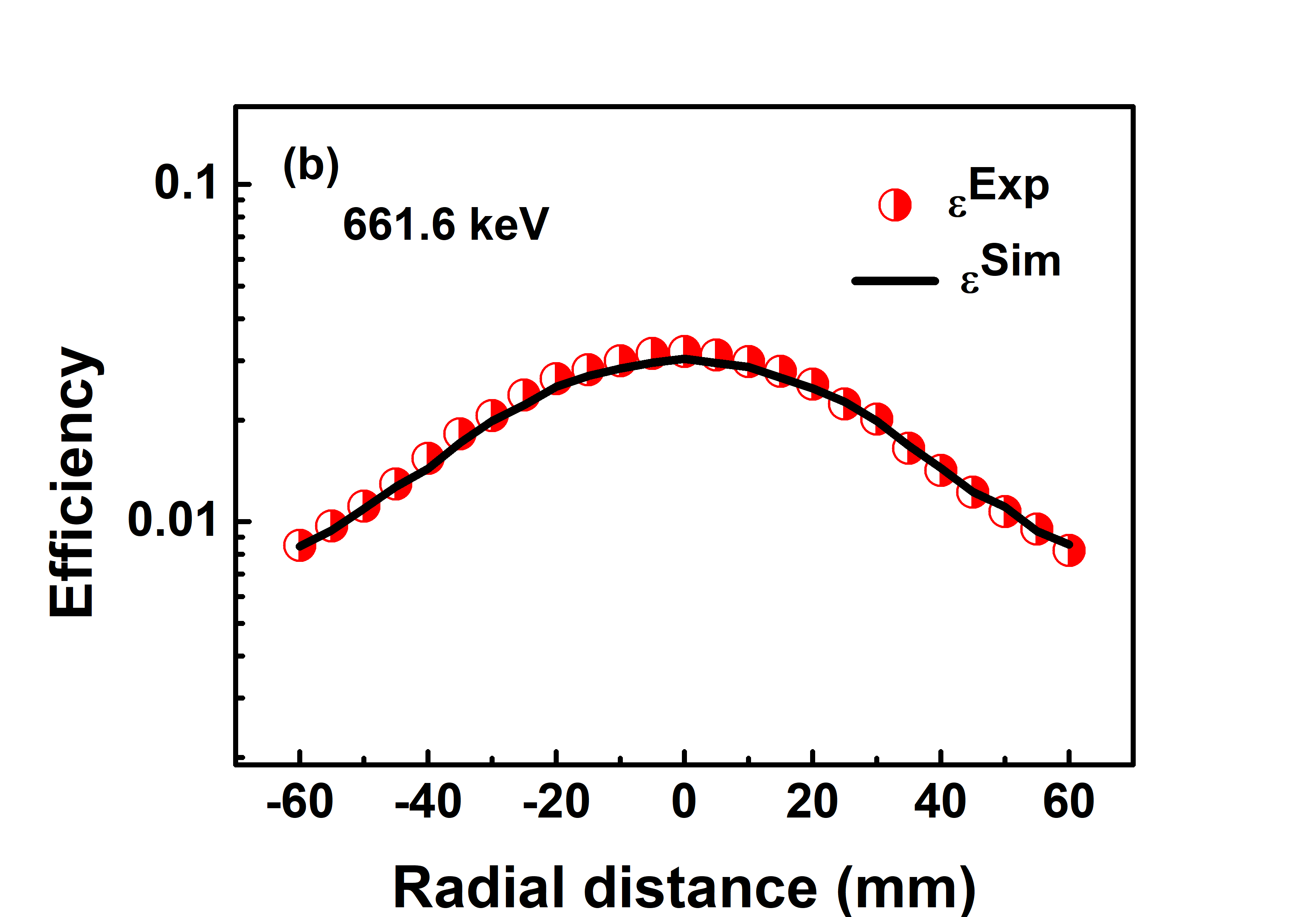}\\
\includegraphics[width=8.5cm,height=5.8cm]{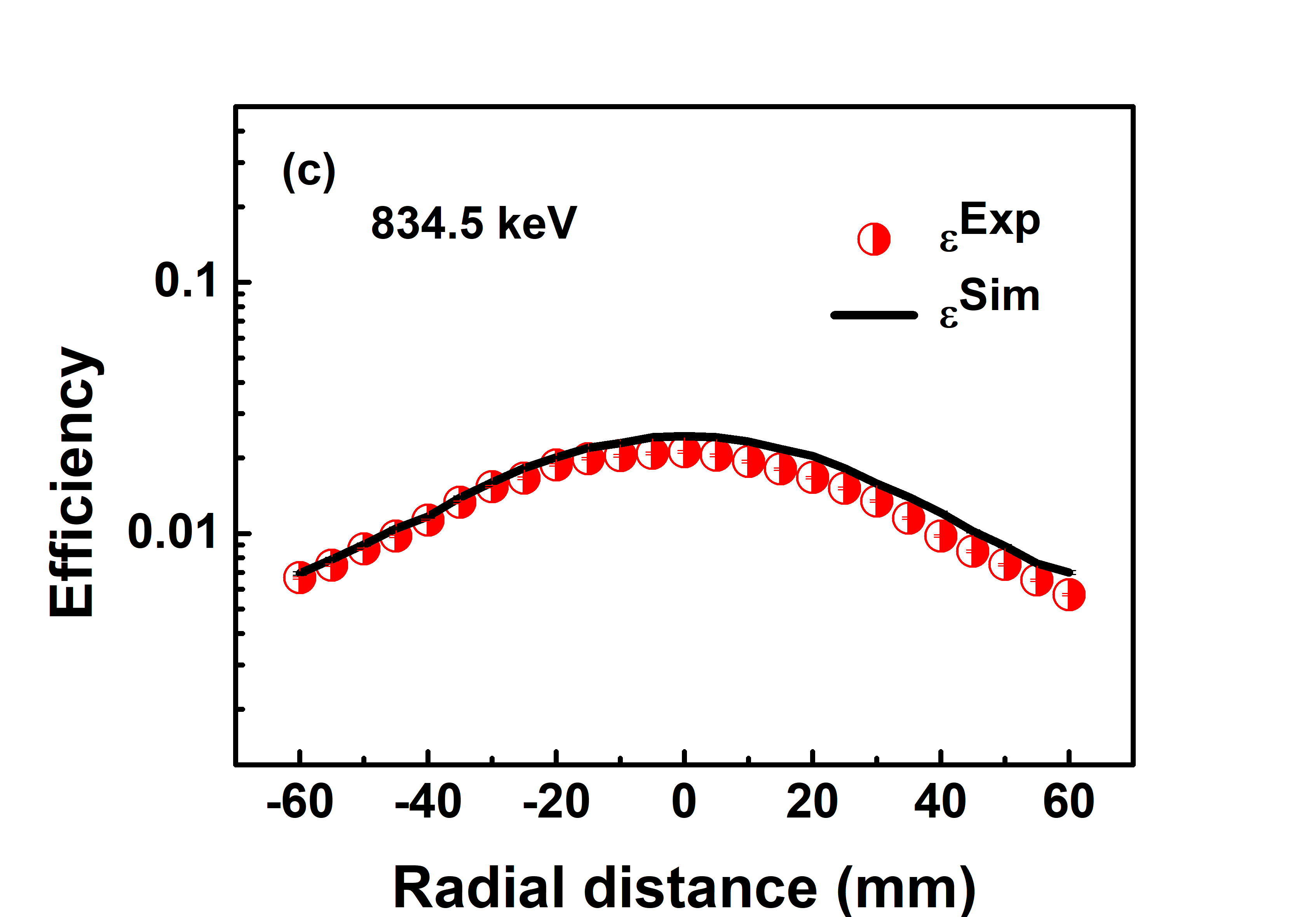}
\includegraphics[width=8.5cm,height=5.8cm]{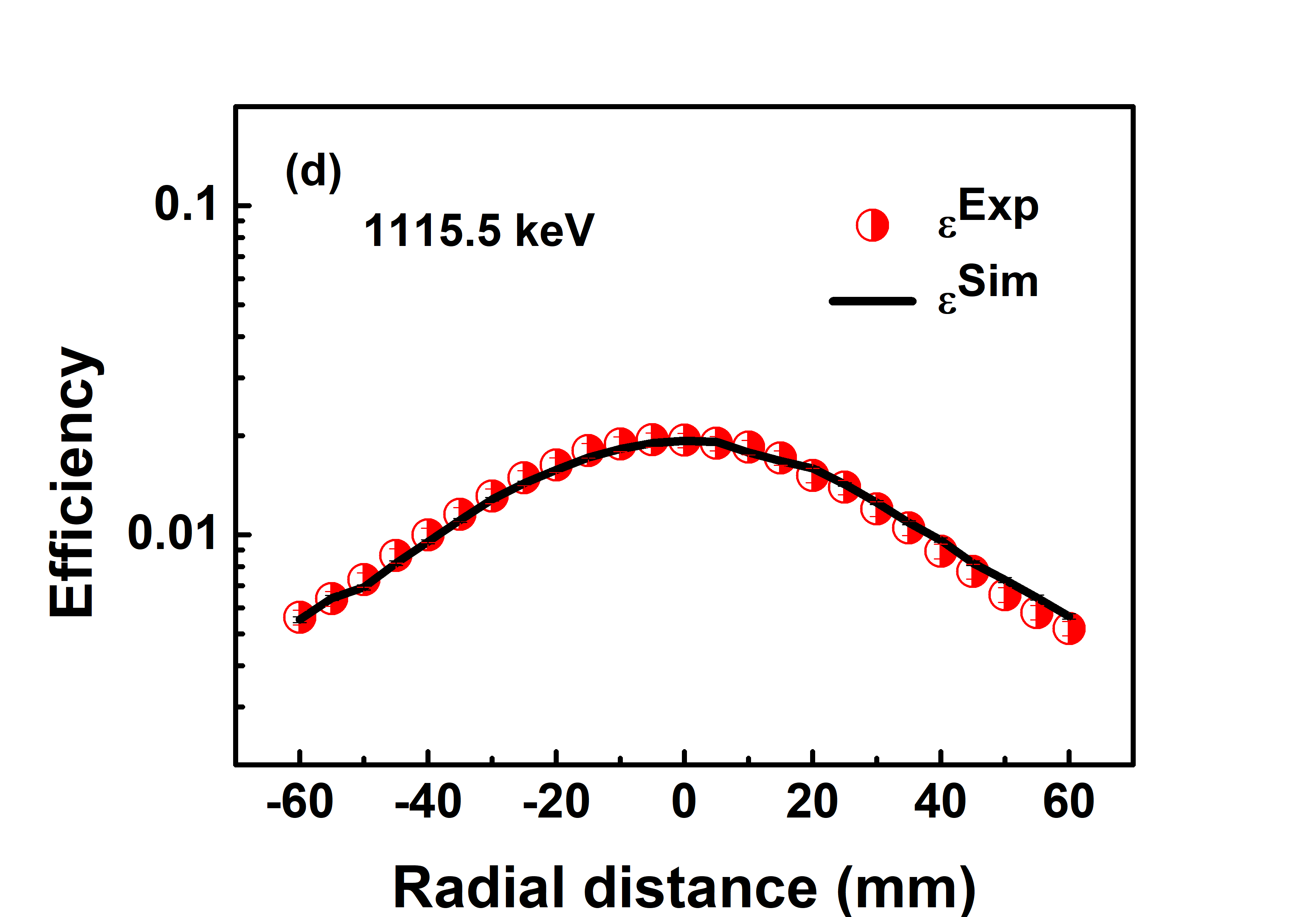}\\
\end{tabular}
\caption{Simulated and experimental absolute photopeak efficiency as a function of radial scan for 122.1~keV, 661.6~keV, 834.5~keV and 1115.5~keV (a), (b), (c) and (d) respectively.}
\label{fig8}
\end{figure}

\begin{figure}[h!]
\centering
\captionsetup{justification=centering}
\begin{tabular}{@{}c@{}}
\includegraphics[width=8.5cm,height=5.9cm]{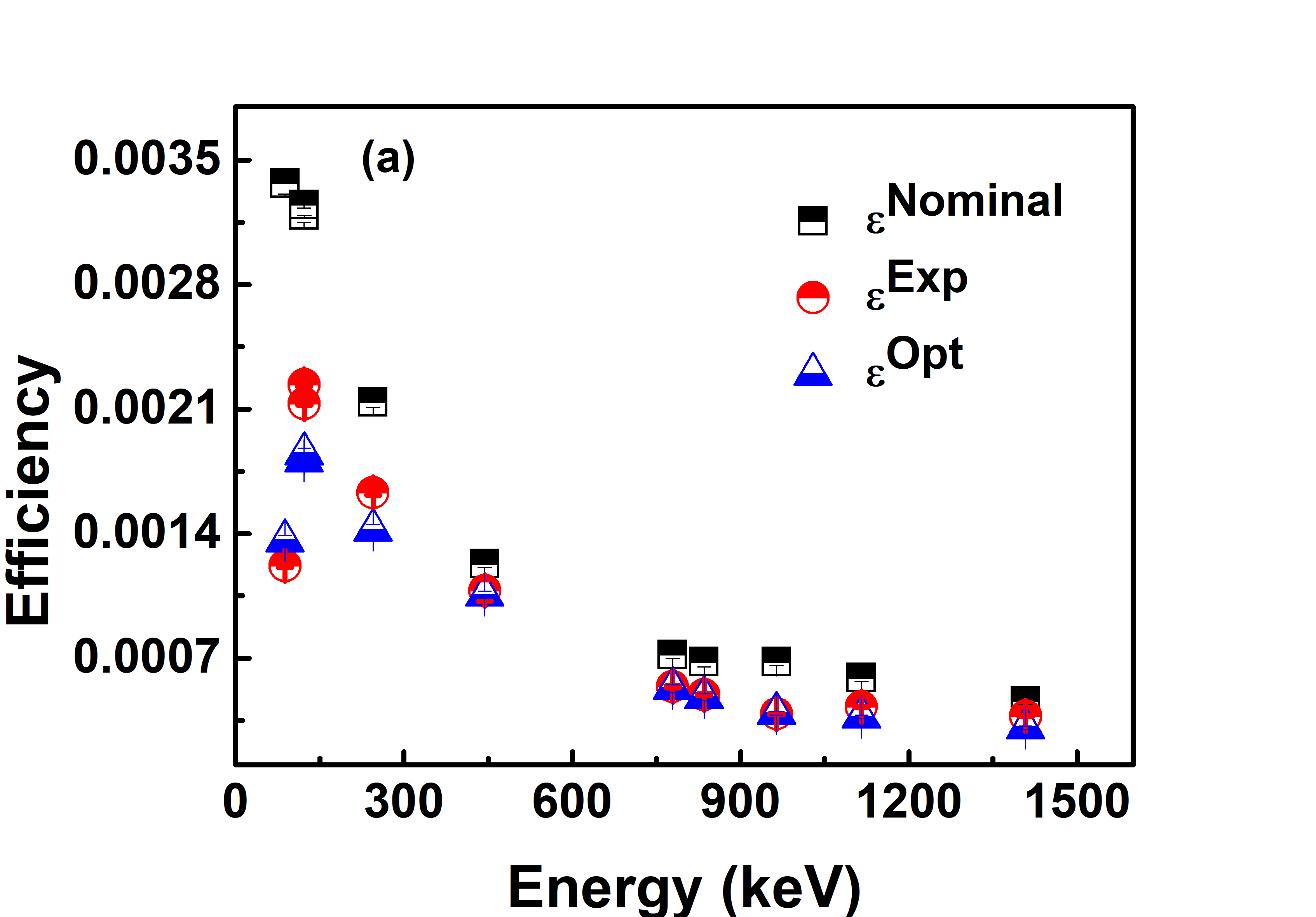}
\includegraphics[width=8.5cm,height=5.9cm]{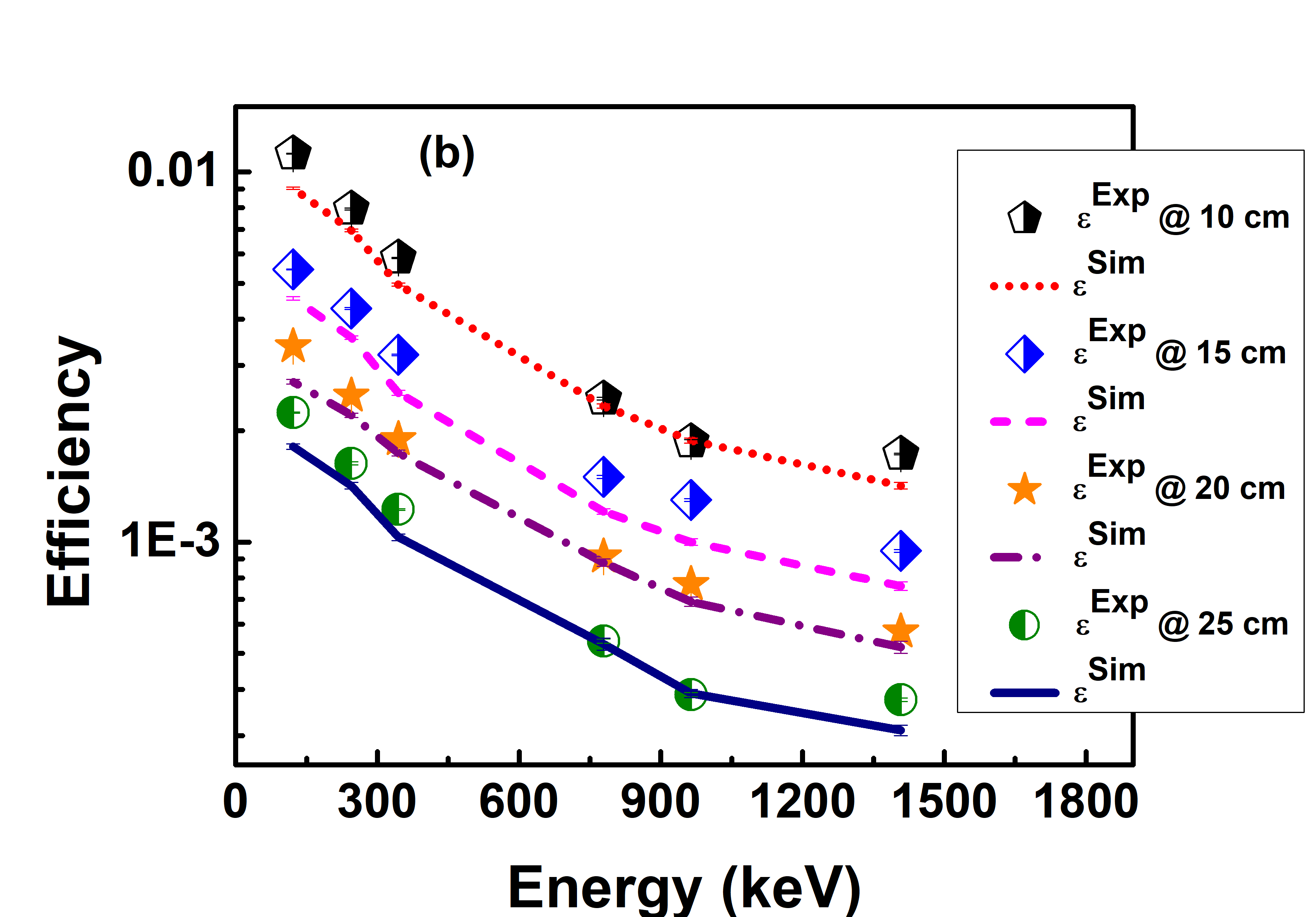}
\end{tabular}
\caption{(a) Comparison of simulated absolute photopeak efficiency using optimized and nominal parameters with the experimental photopeak efficiency. (b) Simulated and experimental absolute photopeak efficiency for $^{152}$Eu at distances 10, 15, 25~cm.} 
\label{fig9}
\end{figure}
Figure~\ref{fig6} to Figure~\ref{fig7} compares distance, radial, and lateral scan data for gamma-rays of various energies with corresponding simulated data using the optimized detector model. Figure~\ref{fig9} illustrates a comparison of the experimental data with simulated efficiencies within the energy range of 88-1408~keV. It is observed that the measured value of 33\% relative efficiency of the detector corresponds to an active volume of about 120 ${\textrm{cm}}^{3}$, which is 11\% smaller than the manufacturer's stated value of 135~${\textrm{cm}}^{3}$. As it is clear from the Figure~\ref{fig10} that {\textrm{$\varepsilon$$^{\textrm{Exp}}$}} and  {\textrm{$\varepsilon$$^{\textrm{Sim}}$}} using the nominal parameters  for E= 122.1-1115.5~keV at d = 5-25~cm produced a significant \textrm{$\sigma$$_{\textrm{TR}}$} of 26.2\%, which was reduced up to 7\% after optimisation.

\begin{figure}[!ht]
\centering
\captionsetup{justification=justified,font=sf,labelfont=bf}
\includegraphics[width=0.58\linewidth,height=0.46\linewidth]{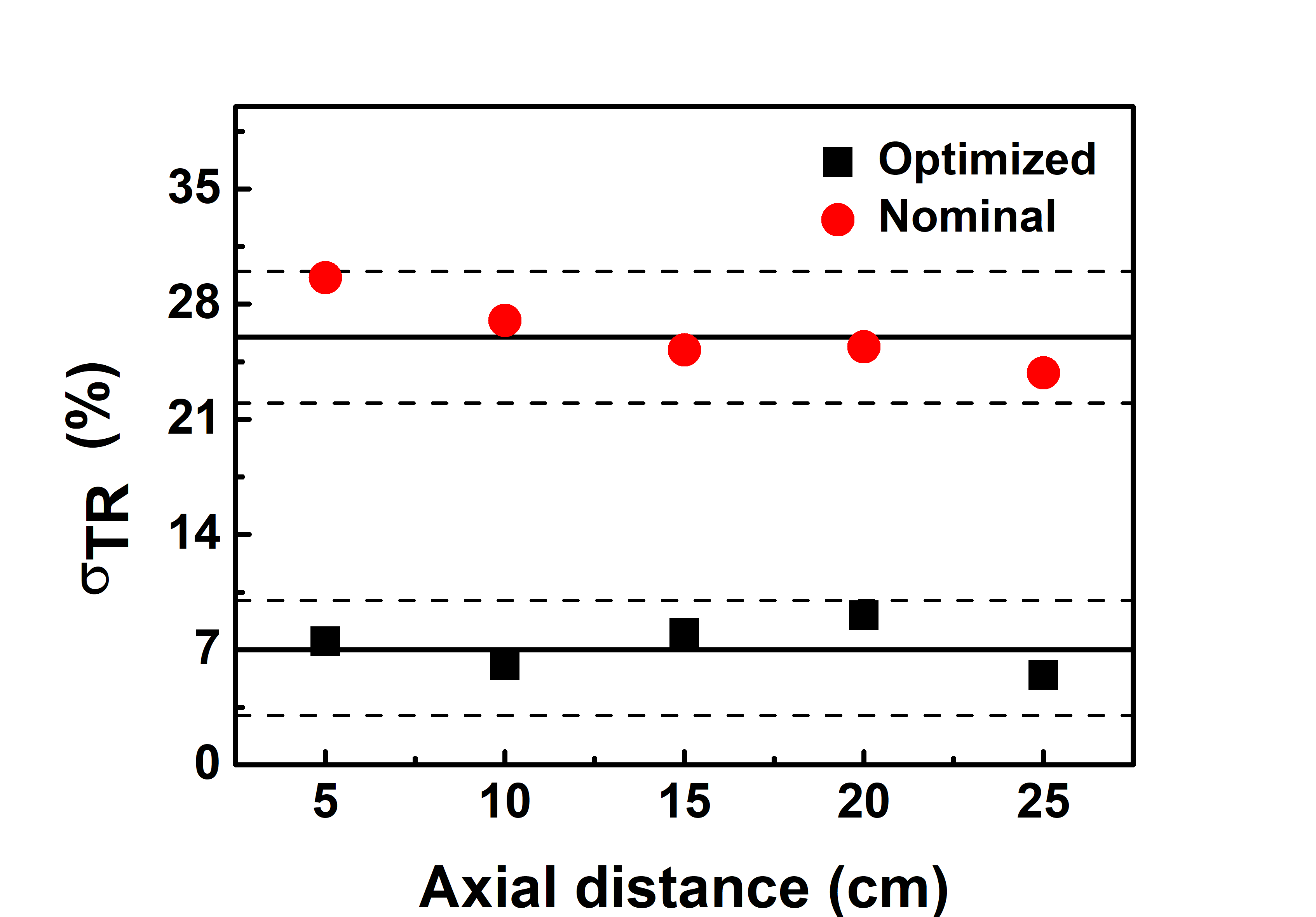}
\caption{Comparison of total relative deviation ($\sigma$$_{TR}$) between nominal and optimized parameters over an energy range of 88-1408~keV for d=5-25~cm.}
\label{fig10}
\end{figure}
\subsection{Radioactivity measurements}\label{Radioactivity measurements}
The ambient background in the laboratory has been measured using an HPGe detector without any shielding at different times since its installation to check for possible changes in the background due to incoming and outgoing materials (for impurity control), daily variations of radon concentration, and aperiodic variations of cosmic rays intensity. From a typical gamma-ray background spectrum recorded for 24~h, about 120 gamma lines were identified, emitted from the uranium and thorium decay chains in the room environment around the detector. The prominent radionuclides present in the spectra are due to the airborne radioactivity of radon and its short-lived decay products, i.e., the intermediate members of radioactive decay series of $^{238}$U and $^{232}$Th. Day-wise investigations of full gamma-ray spectra have shown a considerable variability of the areas of the $^{226}$Ra progeny lines ($^{214}$Pb, $^{214}$Bi) in a range of about 14-18\%. In contrast, the gamma lines of 1460.8\,keV and 2614.5\,keV from $^{40}$K and $^{208}$Tl, respectively, were comparatively constant within the experimental uncertainty. The integral background from energy range of 40\,keV to 2700\,keV is 5.9 $\times$ 10$^{5}$\,/kg/h with a total count rate of 116\,counts/day. Long background measurements were also performed to observe the day-wise variation in the overall background and estimate the statistical fluctuation in the prominent gamma peaks. Most of the variations were attributed to radon dynamics in the laboratory, while the background count rate over the energy region of 40--2700\,keV is found to be similar. The spectrometer exhibits good gain stability and negligible calibration drifts ($<$1~keV) over a long duration of measurements.
A one-day ambient background spectrum of HPGe detector without and with Pb shielding is shown as a representative case in Figure~\ref{fig11}. As a result of lead shielding, the overall background has been significantly reduced from 116 to 2\,counts/day in the energy range of 40-2700\,keV. The background rate for 1460.8~keV ($^{40}$K) and 2614.5~keV ($^{208}$Tl) lines is 634 and 443~counts/day, respectively.
It can be seen that the Pb shield effectively stops most of the gamma-ray from entering the active volume of the detector crystal. At the same time, the reduction ratio of the integration of the count rate depends on the gamma-ray energy. The background reduction achieved has been compared with unshielded data for major peaks and various energy ranges. The ratio is reduced to 1.2\% at 40--600\,keV. In the energy range from 40 to 2700\,keV, the ratio is 1.7\%. The photopeak intensity reduction of major gamma lines visible in the unshielded setup has been compared with a shielded setup as given in Table~\ref{Table5}. For $^{228}$U, the activity levels are reduced to 0.52\% for $^{214}$Bi and $^{214}$Pb. In case of $^{232}$Th, the ratios fluctuate between 0.25\% and 2.46\% for $^{208}$Tl and 0.57\,\% for $^{228}$Ac. Photopeaks of $^{137}$Cs and $^{60}$Co are completely disappeared in the shielded setup. The ratio of $^{40}$K is 1.14\%, and the annihilation peak drops to 20.3\% of its original value without shielding.
\begin{figure} [h]
\centering
\captionsetup{justification=centering}
\begin{tabular}{@{}c@{}}
\includegraphics[width=0.85\linewidth,height=0.65\linewidth]{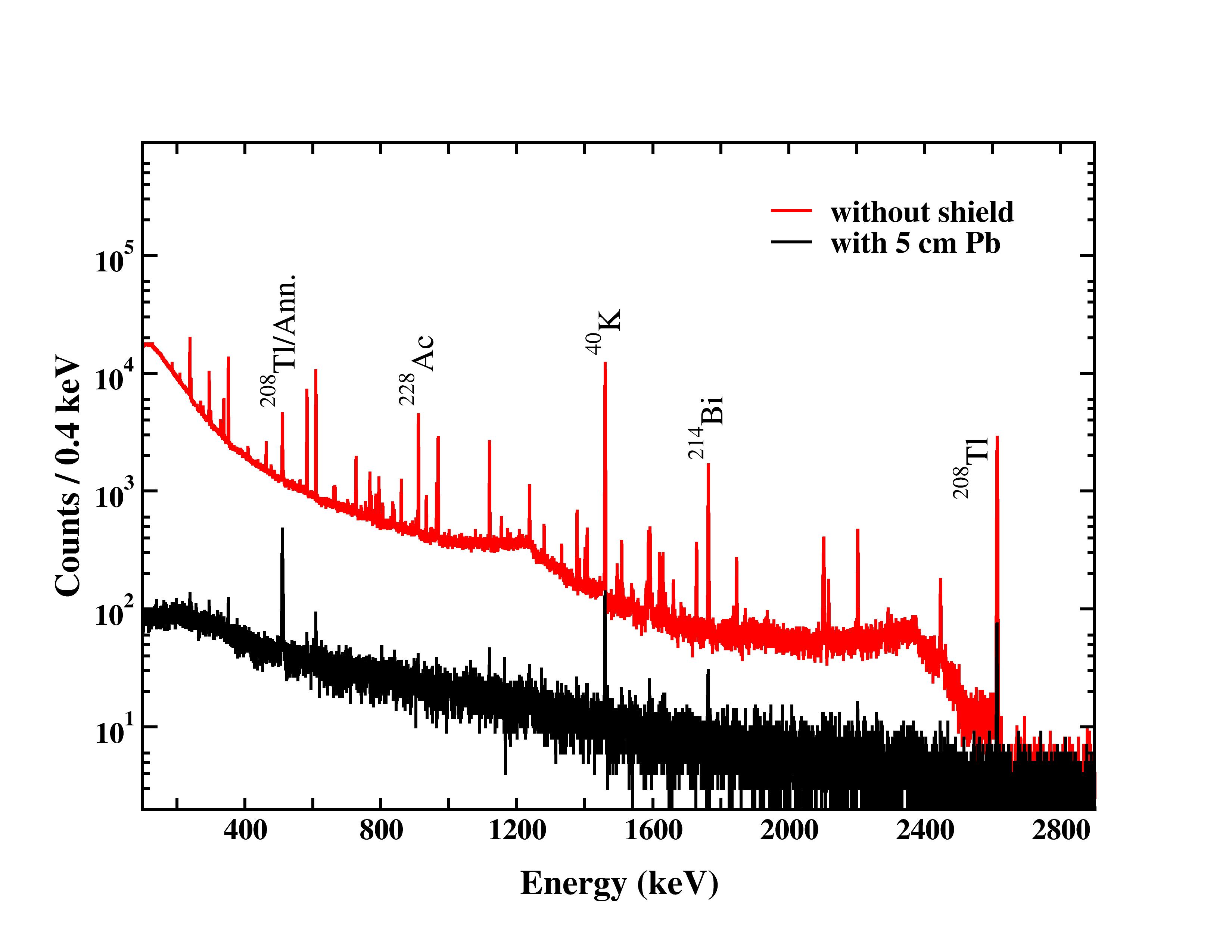}
\end{tabular}
\caption{Ambient gamma-ray background in ILM-0 with (red line) and without (black line) Pb shield (t = 1\,d). Some of the intense gamma-ray lines corresponding to different radionuclides are marked.}
\label{fig11}
\end{figure}
\begin{table}[h!,width=0.684\linewidth,cols=4,pos=h]
\centering
\caption{Observed activity with and without Pb shield along with the reduction factor.}
\begin{tabular*}{\tblwidth}{@{}lllll@{}}
\toprule
Element             & Energy      & Without Shield (X)      & With Shield (Y)          & Reduction factor   \\
                    & (keV)       & (counts/day)        & (counts/day)         &  Y/X(\%)                  \\
\midrule
$^{212}$Pb          & 238.6       & 37812(147)          & 127(47)              & 0.34(3)                     \\
$^{214}$Pb          & 295.3       & 16937(266)          & 86(54)               & 0.51(20)                    \\
$^{214}$Pb          & 351.9       & 31365(329)          & 219(25)              & 0.70(8)                     \\
$^{208}$Tl/Ann.     & 511.0       & 14956(903)          & 3039(110)            & 20.32(12)                   \\
$^{208}$Tl          & 583.2       & 21159(286)          & 54(23)               & 0.25(8)                      \\
$^{214}$Bi          & 609.3       & 32746(349)          & 170(23)              & 0.52(7)                      \\
$^{137}$Cs          & 661.7       & 858(98)             & --                   & --                           \\
$^{214}$Bi          & 806.4       & 1084(139)           & --                   & --                            \\
$^{228}$Ac          & 835.6       & 1356(133)           & --                   & --                            \\
$^{228}$Ac          & 911.2       & 16609(232)          & 94(28)               & 0.57(12)                      \\
$^{214}$Bi          & 1120.3      & 9411(208)           & 106(21)              & 1.13(10)                     \\
$^{60}$Co           & 1173.2      & 654(130)            & --                   & --                            \\
$^{60}$Co           & 1332.5      & 464(87)             & --                   & --                           \\
$^{40}$K            & 1460.8      & 61137(834)          & 700(37)              & 1.14(4)                      \\
$^{214}$Bi          & 1764.5      & 8916(154)           & 199(33)              & 2.23(21)                    \\
$^{208}$Tl          & 2614.5      & 19101(400)          & 469(31)              & 2.64(8)                      \\
\bottomrule
\end{tabular*}
\label{Table5}
 \end{table} 

Ambient background in the shielded setup has been counted for a duration of 49\,d over a span of 3 months. It should be mentioned that the stability of the energy scale was monitored with background gamma-rays such as 1460.8 and 2614.5\,keV, where calibration with standard sources was only sometimes convenient due to heavy passive shielding. In summed gamma-ray background spectra of 49\,d besides gamma-rays emitted by natural radionuclides, another source of background is generated by neutron interactions. The anthropogenic radionuclides $^{137}$Cs and $^{60}$Co were found below detection limits and no contribution from long-lived cosmogenically produced radionuclides, e.g., $^{22}$Na, $^{60}$Co and $^{65}$Zn were observed. In above-ground low-level gamma spectrometry systems, some peaks associated with the activation of germanium and fast neutron scattering in the shielding material occur in the background spectrum. For example, 139.7\,keV gamma line produced via neutron-induced process $^{74}$Ge(n,$\gamma$)$^{75}$Ge. Peaks at 569.7\,keV and 1063.3\,keV from scattering reaction (n,n') on lead were also observed. Another gamma ray at 803.3\,keV is detectable and corresponds to the de-excitation of $^{206}$Pb$^{*}$ which can result from the decay of $^{210}$Po or from fast neutron scattering on lead. 
An important characteristic in low activity measurements is the Minimum detectable activity ($\textrm{A}_{\textrm{D}}$) that defines the least amount of activity in the sample to quantify the radiation level slightly above the unavoidable background. Shorter measurement times and higher background radiation levels would increase the $\textrm{A}_{\textrm{D}}$ and reduce the sensitivity of all detectors. The detector efficiency significantly affects the detection levels. In addition to the detection efficiency, amount of sample, measurement time, and photon emission probability, the $\textrm{A}_{\textrm{D}}$ in gamma spectrometry depends on the background level at a specific energy. These background events mainly come from three primary sources: the sample, the Compton continuum, and natural radioactivity. The $\textrm{A}_{\textrm{D}}$ in Bq/kg at a given gamma-ray energy is calculated using Currie's method~\cite{curie1968} and given by:

\begin{equation}\label{c2:eqn:7}  
\textrm{A}_{\textrm{D}} = \frac{\textrm{N}_{\textrm{D}}}{\textrm{I$_{\gamma}$} \ast \textrm{m} \ast \varepsilon_{\gamma} \ast \textrm{t}}
\end{equation}
Where ${\textrm{N}}_{\textrm{D}}$ = 2.7 + 4.65\,$\sigma_{\textrm{B}}$ is minimum detectable counts and $\sigma_{\textrm{B}}$ is the standard deviation in the background counts, ${\textrm{I}}_{\gamma}$ is the branching ratio of the gamma-ray, $\varepsilon_{\gamma}$ is the photopeak detection efficiency computed using GEANT4 simulation, m is the mass of the sample and t is the counting time. It should be noted that simulated efficiencies ($\varepsilon^{\textrm{Sim}}$) were obtained with the optimized detector model for 10$^6$ events uniformly distributed within the studied sample modeled in GEANT4 similar to the counting geometry kept during the measurement.

\subsection{Measurement of Soil and Rock Samples}\label{Measurement of Soil and Rock Samples}
The sensitivity of the ILM-0 setup was estimated using sub-surface soil (30~cm depth) and rock samples. The soil samples were collected from the Katli village (30.99582~N, 76.534 80~E) in the Ropar region of Punjab (India), and the rock samples were collected from the Aut region of Himachal Pradesh (India). It may be noted that the rock samples were remeasured in the present work to assess the performance of the ILM-0 setup as compared to TiLES measurements reported elsewhere~\cite{thakur2022radiopurity}. The soil samples were oven dried (110$^{\circ}$C), fine powered, and then sieved through a 150~$\mu$m mesh. Each sample was packed and sealed in a cylindrical polypropylene container of $\sim$3.5~cm dia. and $\sim$2.2~cm height and stored for stabilization. The smaller sample geometries are desirable over voluminous sample geometries because the attenuation of gamma-rays within the sample matrix is negligible. A total of 10 soil samples, average mass ⟨m⟩$\sim$20~g were counted in compact geometry for 24~h. The mean soil density ⟨$\rho$⟩, as packed in the container, is estimated to be $\sim$1.3~g/cm$^{3}$ with an overall variation of $\sim$10\%. It has been reported~\cite{SAINI2017} that the Punjab soil is mainly of clay loam type and thus expected to have a similar composition as listed in table~\ref{Table3}. 
\begin{figure}[!ht]
\centering
\captionsetup{justification=justified,font=sf,labelfont=bf}
\includegraphics[width=0.85\linewidth,height=0.65\linewidth]{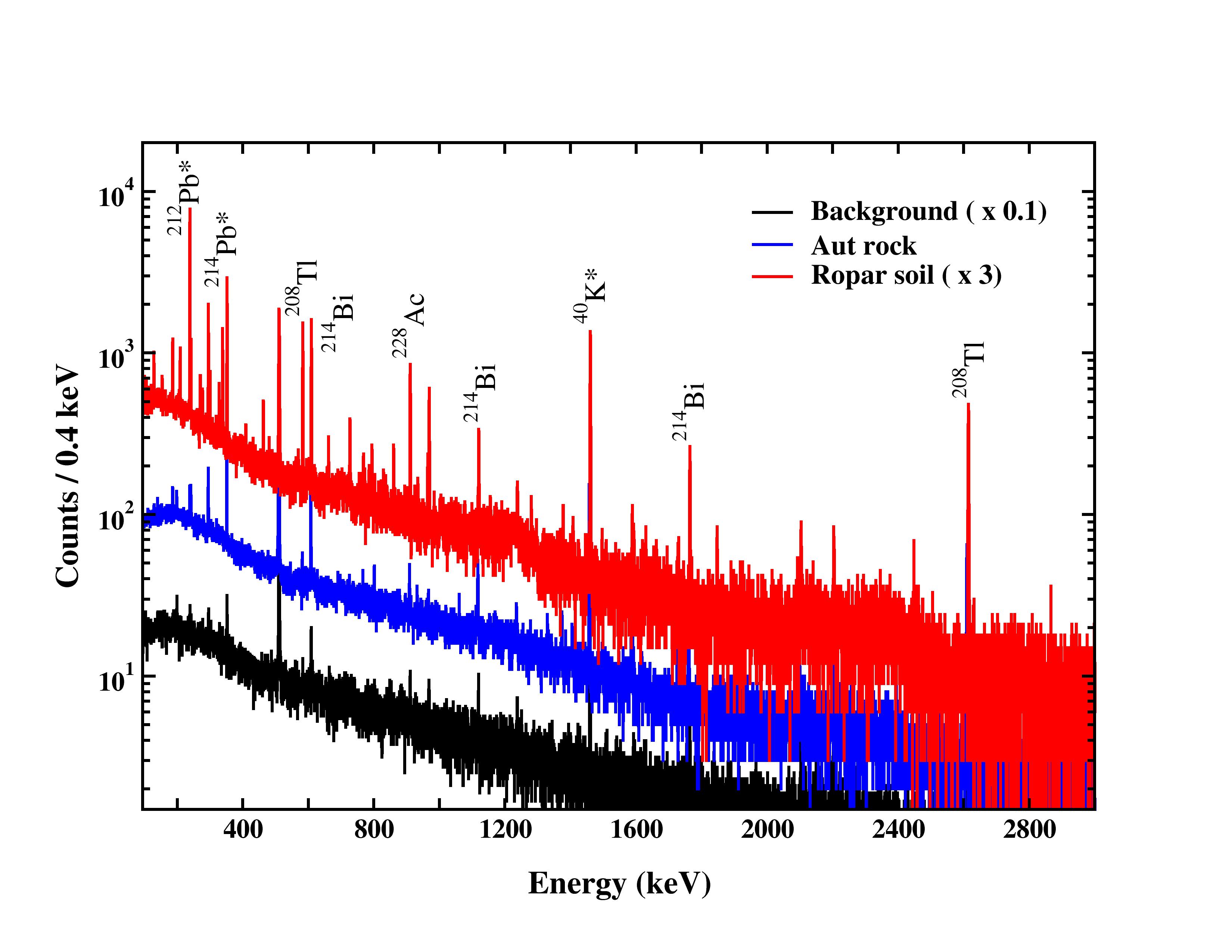}
\caption{A typical gamma-ray spectra of soil sample (red line), Aut rock sample (blue line) and ambient background (black line) without the sample (t=1~d). The soil sample and ambient background are scaled arbitrarily for better visibility. The gamma-rays of interest are indicated (*) in the spectra.}
\label{fig12}
\end{figure}
To enhance the counting efficiency, the soil samples were counted on the front of the detector face and confined within 60\% of the radial extensions to avoid edge effects. The effect of the elemental composition and sample density were taken into account. The average sample mass of ⟨m⟩$\sim$20~g was counted in compact geometry for 24~h. The same geometry was adopted for counting of rock samples. Figure~\ref{fig12} compares typical one-day spectra of ambient background (x0.1), Ropar soil (x3), and Aut rock samples. The lines and notations are self-explanatory. In order to determine the detection limit of ILM-0 for low activity measurements, ${\textrm{A}}_{\textrm{D}}$ for soil matrices were computed from the minimum detectable counts (${\textrm{N}}_{\textrm{D}}$) using the Eq.~\ref{c2:eqn:7}. In close counting geometry, coincident summing affects the observed photopeak yield~\cite{thakur2022radiopurity}. In the present analysis, nuclides in natural radioactive decay chains were considered to be in secular equilibrium, and the gamma-rays with negligible coincidence summing were chosen to estimate activities. ${\textrm{A}}_{\textrm{D}}$ is calculated for several radionuclides often encountered in the environmental samples from which the most dominating ones are shown in Table~\ref{Table6}. MDA values for $^{137}$Cs nuclide have the lowest value and hence determine the nuclide detection limit.
\begin{table}[h!,width=0.5\linewidth,cols=4,pos=h]
\centering
\caption{Estimated sensitivity of the setup.}
\label{Table6}
\begin{tabular*}{\tblwidth}{@{}cccc@{}}
\toprule
Radionuclide                  & Energy      & $\textrm{N}_{\textrm{D}}$       & $\textrm{A}_{\textrm{D}}$  \\
(/Parent)                     & (keV)       & (counts/day)                    & (Bq/kg)                    \\
\toprule
$^{212}$Pb (/$^{232}$Th)      & 238.6       & 65            & 1         \\
$^{214}$Bi (/$^{238}$U)       & 1764.5      & 76            & 25        \\
$^{214}$Pb (/$^{238}$U)       & 295.3       & 85            & 5         \\
$^{228}$Ac (/$^{232}$Th)      & 911.2       & 39            & 4         \\
$^{137}$Cs                    & 661.6       & 3             & 0.1        \\ 
$^{60}$Co                     & 1332.5      & 38            & 2          \\                
$^{40}$K                      & 1460.8      & 120           & 49         \\
\bottomrule
\end{tabular*}
\end{table}
The specific activities were estimated from the measured photopeak yield after background correction, defined as activity per unit mass $\textrm{A}_{\gamma}$ corresponding to a given transition of the radionuclide was determined using,
\begin{equation}\label{c2:eqn:8}  
{\textrm{A}}_{\gamma} = \frac{{\textrm{N}}_{\gamma}}{\textrm{I$_{\gamma}$} \ast \textrm{m} \ast \varepsilon_{\gamma} \ast \textrm{t}},
\end{equation}
where $\textrm{N}_{\gamma}$ is the net observed counts in the photopeak after correcting for the ambient background. The uncertainty in the specific activity includes the error in efficiency and peak fitting. Most of the gamma-rays were visible from the uranium and thorium decay chains. However, only those gamma-rays that could be unambiguously assigned to a particular nuclide were considered for further analysis. Therefore, correction factors due to sample self-absorption and coincidence summing can be reasonably neglected. For $^{238}$U single gamma line of $^{214}$Pb at 295.3\,keV, for $^{232}$Th gamma line of $^{212}$Pb at 238.6\,keV and $^{40}$K 1460.8\,keV were selected for the analysis. The observed specific activity of primordial radionuclides in the soil samples was found to range between 32–67, 66–107, and 590–860 with a mean specific activity of 50, 85, and 670~Bq/kg for $^{238}$U, $^{232}$Th, and $^{40}$K, respectively. The measured activity of $^{238}$U and $^{232}$Th showed a consistent distribution of primordial radionuclides among all the soil samples, while a somewhat large scatter is observed in $^{40}$K data.
In the case of rock sample measurements, the measured specific activity of $^{238}$U in Aut rock is 7 (1)~Bq/kg, which is similar in comparison to the value reported in Ref~\cite{thakur2022radiopurity}, while for $^{232}$Th and $^{40}$K no measurable activity could be observed above the ambient background at the present experimental sensitivity. 
To estimate the activity concentration in rock samples, reducing the overall background by an order of magnitude is necessary, as seen in Figure~\ref{fig12}. The existing setup needs augmentation with thicker low activity lead/copper shields to improve the detection limits. As discussed earlier, above-ground laboratories are mostly dominated by muon-induced interactions; therefore, deploying additional cosmic muon veto systems are desirable to improve the sensitivity with an overall background reduction of about 50\% in the present setup.

\section{Summary and Conclusions} \label{Summary and Conclusions}
In summary, a moderately shielded low background counting facility has been set up at IIT Ropar to investigate environmental radioactivity and rare decays. The GEANT4 simulations employing Monte Carlo procedures have been performed to determine the photopeak efficiency of the HPGe detector, and an effective detector model has been obtained. Various parameters, such as; crystal radius (R), crystal length (L), front gap (g), dead layers (t$_{d}$, t$_{b}$, and t$_{s}$) have been compared and optimized through the lateral, radial and distance scanning measurements. The relative deviations between simulated and experimental efficiencies were found to be $\sim$7\% within the photon energy range of 80.9--1408~keV, indicating as high as $\sim$93\% confidence level in the comparison between experimental characterization and simulation. In order to assess the performance of the setup, the trace concentration of radioactive elements has been measured in soil samples from the agricultural test site in Ropar with ILM-0. The studied site has mean specific activity of 50, 85, and 670~Bq/kg for $^{238}$U, $^{232}$Th, and $^{40}$K, respectively. While $^{238}$U and $^{232}$Th activities among the samples are consistent within overall uncertainty, $^{40}$K shows a somewhat more significant variation. The Aut rock measurements have shown specific activity of $^{238}$U to be 7 (1) Bq/kg, while for $^{232}$Th, and $^{40}$K furthur measures to improve signal-to-noise ratio are essential. Efforts are underway to augment the setup with low activity lead shields and plastic veto detectors to improve the sensitivity for studies relevant to rare decay searches. Measurements of the studied site with rich datasets employing two detector configurations are proposed.

\section{Acknowledgments} \label{Acknowledgments}
The authors thank Prof. R. G. Pillay and Prof. Vandana Nanal for their suggestions during the development of the setup. The Indian Institute of Technology Ropar acknowledged for an ISIRD grant for setting up ILM-0 for rare decay physics. One of the authors, Swati Thakur thanks the Ministry of Education (MoE), Government of India, for the doctoral fellowship, and iHub - AWaDH, a Technology Innovation Hub established by the Department of Science $\&$ Technology, Government of India, in the framework of the National Mission on Interdisciplinary Cyber-Physical Systems (NM - ICPS), for financial support to execute this work. Soni Devi and Katyayni Tiwari thank the doctoral fellowships received from the University Grant Commission and DST - INSPIRE, respectively. 

\bibliographystyle{elsarticle-num}

\bibliography{bibliography}

\end{document}